\documentclass[aps,preprintnumbers,nofootinbib,onecolumn,superscriptaddress,notitlepage]{revtex4-1}

\usepackage{graphicx}
\usepackage[dvipsnames]{xcolor}
\usepackage{caption}
\usepackage[shortlabels]{enumitem}
\usepackage{mathtools}
\usepackage{fancyhdr}
\usepackage{nicefrac}
\usepackage{bbm}
\usepackage{soul}
\makeatletter

\usepackage{upgreek}
\newcommand{\changefont}{
    \fontsize{9}{11}\selectfont
}
 \usepackage{makecell}
 \usepackage{placeins}
  \usepackage{afterpage}
\usepackage{mathrsfs}

\usepackage{graphics}
\usepackage{amsfonts,amssymb,amsmath}            
\usepackage{ifthen} 
\usepackage{amsthm, thm-restate}
\usepackage{dsfont}
\usepackage{float}
\usepackage{MnSymbol}
\usepackage{comment}

\usepackage[caption=false]{subfig}
\PassOptionsToPackage{hyphens}{url}
\usepackage[hyperfootnotes=false]{hyperref}
\usepackage{xcolor}
\definecolor{mylinkcolor}{rgb}{0,0,0.7}
\hypersetup{unicode=true,
  bookmarksnumbered=false,bookmarksopen=false,bookmarksopenlevel=1, 
  breaklinks=true,pdfborder={0 0 0},colorlinks=true}
\hypersetup{
  anchorcolor=mylinkcolor,citecolor=mylinkcolor, 
  filecolor=mylinkcolor,linkcolor=mylinkcolor, 
  menucolor=mylinkcolor,runcolor=mylinkcolor, 
  urlcolor=mylinkcolor}

\usepackage{tikz}
\usetikzlibrary{arrows}
\usetikzlibrary{arrows.meta}
\usepackage{rotating, xcolor}
\usetikzlibrary{shapes}
\usetikzlibrary{hobby}
\usetikzlibrary{decorations.markings}
\usetikzlibrary{arrows.meta}
\usetikzlibrary{snakes} 
  
\usepackage[nodayofweek]{datetime}
\newcommand{\ket}[1]{| #1 \rangle}
\newcommand{\bra}[1]{\langle #1 |}
\newcommand{\kket}[1]{| {#1} \rangle\rangle} 
\newcommand{\bbra}[1]{\langle\langle {#1} |} 
\newcommand{\braket}[2]{\langle #1|#2\rangle}
\newcommand{\ketbra}[2]{|#1\rangle\!\langle#2|}
\newcommand{\proj}[1]{|#1\rangle\!\langle#1|}

\newcommand{\tr}{{\rm Tr}}

\theoremstyle{plain}

\theoremstyle{definition}

\begin{document}

\title{
Time delocalization and causality across temporal quantum reference frames}

\author{Veronika Baumann}
\email{veronika.baumann@oeaw.ac.at}

\affiliation{Institute for Quantum Optics and Quantum Information (IQOQI),
\\Austrian Academy of Sciences, 1090 Vienna, Austria}
\affiliation{Atominstitut, TU Wien, 1020 Vienna, Austria}

\author{Maximilian P. E. Lock}
\email{maximilian.paul.lock@tuwien.ac.at}
\affiliation{Atominstitut, TU Wien, 1020 Vienna, Austria}
\affiliation{Institute for Quantum Optics and Quantum Information (IQOQI),
\\Austrian Academy of Sciences, 1090 Vienna, Austria}

\begin{abstract}
 \noindent In relational quantum dynamics, evolution emerges via the correlations between some system of interest and a clock system, which plays the role of a temporal reference frame. Their combined state satisfies a Wheeler-de Witt-like constraint equation, and therefore  does not evolve, leading to a ``block universe'' picture. Here we investigate the interplay of two aspects, namely temporal localization and causal relations, when comparing emergent dynamics with respect to different choices of clock. We first explore the extent to which two clocks can agree on the temporal localization of events. Then, focussing on the operational notion of causality, we require a clearly defined notion of interventions, i.e. quantum operations, and consider two different approaches to modeling these operations within relational dynamics. The first considers their application via the choice of solutions to the constraint equation, i.e.~the choice of which ``history'' is considered. The second approach incorporates the operations into the constraint equation itself and thereby into its solutions, giving a dynamical picture of the interventions. From the perspective of a single clock, both approaches allow for a notion of operational causality in relational dynamics. However, for multiple clocks, only the second approach gives a consistent picture regarding causal relations, while necessarily manifesting some degree of temporal delocalization between frames. Moreover, this second approach, when considering certain cases of temporal delocalization, naturally describes scenarios with indefinite causal order, a well-known quantum feature of operational causality.
\end{abstract}

\maketitle

\tableofcontents

\section{Introduction}
\label{sec:introduction}

By describing the properties of one quantum system relative to another, treated as a quantum reference frame (QRF), we can construct a relational picture of physics that minimizes reliance on idealized background structures such as classical spacetime coordinates. When applied to time, this perspective offers a compelling route for addressing the ``problem of time'' in canonical quantum gravity~\cite{rovelli1990quantum,Isham1993,kucharTimeInterpretationsQuantum2011a}. More broadly, QRFs provide a platform for constructing theories in which space and time emerge from the relations among physical systems, breathing new life into the longstanding philosophical debate between the relational and the substantivalist understandings of spacetime~\cite{teller1991substance,anderson2017problem,adlam2025we}.

However, the pursuit of this program raises significant challenges. Unlike classical reference frames, different choices of QRF can lead to radically different descriptions of physical systems, which in the context of temporal QRFs is known as the ``multiple clock problem''~\cite{kucharTimeInterpretationsQuantum2011a}. Crucially, different choices of clock lead to different tensor factorizations of the Hilbert space, and in some cases to the loss of distinction between subsystems~\cite{ali2022quantum}. While one may adopt a fully clock-neutral perspective and interpret these problems as simply features of a relational world, the operational comparison of events in different frames remains an open problem~\cite{kabel2025quantum}, particularly since the joint temporal localizability of events required for such a comparison seems to be ruled out by the gravitational interaction~\cite{castro2020quantum}. 
Indeed, this issue becomes especially acute in a gravitational context, as spacetime geometry is (up to a conformal factor) determined by the causal relations of the events comprising that spacetime~\cite{malament1977class}, leading to the hypothesis that quantum gravity extends quantum indeterminacy to causal structure itself~\cite{hardy2005probability,hardy2007towards,chiribella2013quantum,oreshkov2012quantum,surya2019causal}. How, then, can causal structure be understood within in a fully relational quantum framework, and how can it be compared across reference frames, particularly given the lack of joint localizability mentioned above?

A subtlety arises here due to inequivalent but overlapping notions of causality. Relativistic causality is associated with the situation of events in the past and future light-cones of each other, expressed in the context of quantum field theory via microcausality conditions, i.e. the commutation or anti-commutation of observables at spacelike separations. In contrast, information-theoretic approaches define causality in operational terms via the correlations between inputs and outputs of a given process (e.g. the dynamical evolution of some system), for example by considering the effect of interventions on a given input to the process~\cite{pearl2009causality,allen2017quantum,barrett2019quantum}. The study of indefinite causal order is situated within a similarly operational setting~\cite{chiribella2013quantum,oreshkov2012quantum}. The difference between the two notions of causality is highlighted by the fact that a processes exhibiting indefinite causal order in the operational sense can be embedded in a fixed spacetime~\cite{vilasini2024fundamental,vilasini2024embedding}.

Some first steps have been made in the study of causal relations in a relational setting, such as in recent work exploring how the identification of spacetime points becomes frame-dependent in quantum gravity~\cite{kabel2025quantum}, and how the definite or indefinite character of causal order, defined through proper-time differences between worldline intersections, remains invariant under changes of quantum reference frame~\cite{de2025indefinite}. In the context of field theory, a relational description can provide an alternative to~\cite{hohn2024matter}, or a means to recover~\cite{ben2025quantum,hoehn2025fighting}, relativistic microcausality conditions.

Here, we will focus on the operational notion of causality, specifically its embedding into relational dynamics, and its intersection with the problem of temporal localization. We investigate the extent to which different reference frames can agree about the temporal localization of an event, and how they can disagree about the direction of time, the ordering of operations, and the causal interpretation of interventions. We adopt the relational picture afforded by the Page-Wootters formalism~\cite{PageWootters1983,woottersTimeReplacedQuantum1984,pageClockTimeEntropy1994}, whose equivalence to other formalisms for relational dynamics has been explored in~\cite{Trinity2021,hohn2021equivalence,chataignier2026relational}. We study how an intervention can be modeled in this picture, and the necessary role of temporal localization as well as its implications for causal explanations across temporal reference frames.
We begin by briefly reviewing the Page-Wootters formalism in Sec.~\ref{sec:PWformalism}. In Sec.~\ref{sec:clock-synch} we discuss in more detail the concept of time delocalization when dealing with multiple temporal reference frames and in Sec.~\ref{sec:time-reversal} we examine whether the order of operations can, nevertheless, be reference-frame-dependent. In Sec.~\ref{sec:OperationalCausality&PW} we review the operational approach to causal relations before presenting a naive attempt to incorporate it into relational dynamics. The latter turns out to be problematic when comparing the perspective of different clocks, as we show in Sec.~\ref{ssec:OperationalCausalityIssues}. Consequently we discuss how clock-system interaction terms in the constraint operator can model timed interventions in the Page-Wootters formalism in Sec.~\ref{sec:interventions&PW}. This not only allows us to fully incorporate standard operational causality into relational dynamics but also for describing instances of indefinite causal order. Finally we discuss our results in Sec.~\ref{sec:Discussion}.

\section{The Page-Wootters formalism}
\label{sec:PWformalism}

The Page-Wootters formalism describes the evolution of one or more quantum systems in relation to another one, which acts as a temporal reference frame i.e. a clock~\cite{PageWootters1983,woottersTimeReplacedQuantum1984,pageClockTimeEntropy1994}. Dynamics emerge from a Wheeler-DeWitt-like constraint equation as we explain below. We then discuss important aspects of considering multiple clock systems with respect to which relational dynamics can be obtained, and how the evolution depends on which clock one chooses as the reference frame.

\subsection{Evolution relative to a single clock}
\label{ssec:singelclockPW}

The Page-Wootters formalism was introduced in response to the problem of time in canonical quantum gravity~\cite{Isham1993,kucharTimeInterpretationsQuantum2011a}, whereby the wavefunction of the universe is stationary. Given a so-called kinematical Hilbert space $\mathcal{K}$, and a constraint operator $\hat{C}$ (identified with the Hamiltonian) defined thereupon, the \emph{physical states} are the solutions of a constraint i.e. Wheeler-DeWitt equation~\cite{dewitt1967quantum}:
\begin{align}
    \label{eq:constraint equation}
    \hat{C}\kket{\Psi}=0 .
\end{align}
We denote the physical Hilbert space, comprised of the set of all such $\kket{\Psi}$, by $\mathcal{H}_{\rm phys}$. Let us for now consider a simple toy setting, namely where the kinematical Hilbert space $\mathcal{K}$ is composed of the Hilbert spaces of some system $S$, whose evolution one is interested in, and that of a clock $C$, i.e. $\mathcal{K}=\mathcal{H}_C\otimes\mathcal{H}_S$. Moreover, consider a constraint Hamiltonian which separates into clock and system terms $\hat{H}_C$ and $\hat{H}_S$ respectively:
\begin{align}
    \label{eq:Schrödingerconstraint}
    \hat{C}=\hat{H}_C+\hat{H}_S,
\end{align}
where, for ease of notation, we neglect tensor factors of identity throughout this work, except where their inclusion is illustrative. For so-called \emph{ideal clocks} one can define a self-adjoint operator $\hat{T}$ via $[\hat{H}_C,\hat{T}]=i$, which can then be expressed in terms of its (improper) eigenstates $\ket{t}$\footnote{More generally, outside of the ideal setting, one defines time states and their corresponding non-projective POVM via a covariance condition~\cite{smith2020quantum,Trinity2021}. While we will later relax some of the assumptions introduced in this section, we will maintain the assumption of ideal clocks throughout this work, unless stated otherwise.}
\begin{align}
    \label{eq:Toperator}
    \hat{T}= \int_{\mathds{R}} dt \, t \, \proj{t}_C ,
\end{align}
where $t$ is the parameter time of the clock $C$, and where $e^{-i\hat{H}_{C}t'}\ket{t}=\ket{t+t'}$. Since the vectors $\ket{t}$ form a basis we can expand the solutions to the constraint equation as follows
\begin{align}
    \label{eq:t-expansionPsi}
    \kket{\Psi}= \int_{\mathds{R}} dt \, \ket{t}_C\ket{\psi_{|C}(t)}_S,
\end{align}
where $\ket{\psi_{|C}(\tau)}\coloneqq\langle\tau\kket{\Psi}$ are the \emph{conditional states of the system} corresponding to the clock reading $t=\tau$. Throughout this work, we will use $t$ as the label for elements of the basis of time states, and $\tau$ in the case where we refer to a specific element of this basis, e.g. when describing conditional states. We use the subscript ``$|C$'' to denote that a quantity is being described relative to $C$. While the solutions to the constraint equation are static, the conditional system states evolve relative to the clock times given by $\hat{T}$. For a constraint Hamiltonian in the form of Eq.~\eqref{eq:Schrödingerconstraint}, this relational evolution is given by the Schrödinger equation. To see this, first note that we can find solutions to the constraint equation by averaging over the group generated by $\hat{C}$:
\begin{align}
\label{eq:physical_state_from_deltaC}
    \kket{\Psi}=\delta(\hat{C})\ket{\phi}\equiv \frac{1}{2\pi} \int_{\mathds{R}} ds\, e^{-is\hat{C}}\ket{\phi},
\end{align}
where $\ket{\phi}$ is an arbitrary state from the kinematical Hilbert space $\mathcal{K}$ and $\delta(\hat{C})$ is an (improper) projector onto the physical Hilbert space $\mathcal{H}_{\rm phys}$. Note that, when constructing physical states in this manner, the inner product on $\mathcal{H}_{\rm phys}$ can be written as  $\braket{\langle\Psi}{\Psi\rangle}_\mathrm{phys}=\bra{\phi}\delta(\hat{C})\ket{\phi}=\braket{\phi}{\Psi\rangle}$ (see~\cite{Trinity2021} Sec.~IV~A for a detailed discussion). Further note that $\delta(\hat{C})$ is many-to-one, i.e. there are multiple $\ket{\phi}\in\mathcal{K}$ that give rise to any given state $\kket{\Psi}\in\mathcal{H}_\mathrm{phys}$.

We will now examine how the representation of kinematical states in a time basis relates to the form of the associated physical states. To wit, using the eigenstates of $\hat{T}$ we can write an abitrary kinematical state as  $\ket{\phi}=\int_{\mathds{R}} dt\,\sumint_{\lambda} \varphi(t,\lambda)\ket{t}_C\ket{\lambda}_S$, where $\lambda$ labels an arbitrary basis of $\mathcal{H}_S$ and $\sumint$ indicates either a sum or an integral according to the discrete or continuous nature of this basis. 
The constraint operator in Eq.~\eqref{eq:Schrödingerconstraint} then leads to:
\begin{align}
   \label{eq:WPgeneral1} 
   \kket{\Psi} &=  
   \frac{1}{2\pi}\int_{\mathds{R}} ds\, dt\, \ket{t+s}_C \sumint_{\;\lambda} \varphi(t,\lambda) e^{-is \hat{H}_S}\ket{\lambda}_S, \\
   \label{eq:WPgeneral2} 
   \ket{\psi_{|C}(\tau)}= \braket{\tau}{\Psi}\rangle &= e^{-i\tau\hat{H}_S}  \sumint_{\;\lambda}\int_{\mathds{R}} dt \,\varphi(t,\lambda) e^{it\hat{H}_S}\ket{\lambda}_S \quad \text{and} \quad i\frac{d}{d\tau}\ket{\psi_{|C}(\tau)}= \hat{H}_S \ket{\psi_{|C}(\tau)},
\end{align}
where the \emph{conditional states} $\ket{\psi_{|C}(\tau)}$ belong to a Hilbert space which we denote $\mathcal{H}_{\vert C}$. Note that, despite the possibility of correlations between clock and system in the kinematical space, Eq.~\eqref{eq:physical_state_from_deltaC} yields the same physical state as in Eq.~\eqref{eq:WPgeneral1} when we take a (non-normalizable) kinematical state of the form~$\ket{\phi}=\ket{t_0}_C\ket{\phi_0}_S$ with $t_{0}$ arbitrary.\footnote{This follows from identifying $\ket{\phi_{0}}=e^{-it_{0}\hat{H}_{S}}\sumint_{\;\lambda}\int_{\mathds{R}} dt \,\varphi(t,\lambda) e^{it\hat{H}_S}\ket{\lambda}_S$.}  As long as $\ket{\phi_0}_{S}$ is normalized in $\mathcal{H}_{S}$, this gives normalized physical and conditional states. More concretely, Eqs.~\eqref{eq:WPgeneral1} and~\eqref{eq:WPgeneral2} become
\begin{align}
\label{eq:WPsimple}
    &\kket{\Psi}= \frac{1}{2\pi} \int_{\mathds{R}} ds\, \ket{t_0+s}_C e^{-is\hat{H}_S}\ket{\phi_0}_S,\\
    &\ket{\psi_{|C}(\tau)}= e^{-i(\tau-t_0)\hat{H}_S}\ket{\phi_0}_S.
\end{align}
In the following sections we will use the above form of kinematical state to generate examples of physical (and therefore conditional states).\\

More generally, one may consider a constraint operator which couples the clock and the system:
\begin{align}
    \label{eq:constraint_int}
    \hat{C}=\hat{H}_{C}+ \hat{H}_{S}+\hat{H}_{\rm int}.
\end{align}
We can again express the physical state as in Eq.~\eqref{eq:t-expansionPsi} although, the conditional system states in general no longer evolve according to the Schrödinger equation. As shown in~\cite{Smith2019} if $\bbra{\Psi}[\proj{t}\otimes \mathds{1}_S,\hat{H}_{\rm int}]\kket{\Psi}=0$ the system $S$ evolves according to a modified Schrödinger equation instead, which still corresponds to unitary relational dynamics.

\subsection{Relational dynamics with multiple clocks}
\label{ssec:multiclocksPW}

We now discuss the Page-Wootters formalism with \emph{multiple clocks}, i.e the scenario where there are multiple quantum systems that we may choose as temporal reference frames, and review how to change between these different frames. We therefore consider the kinematical Hilbert space $\mathcal{K}=\bigotimes_i \mathcal{H}_{C_i}\otimes \mathcal{H}_S$ and constraint operators of the form 
\begin{align}
    \label{eq:constraint_general}
    \hat{C}=\sum_i \hat{H}_{C_i}+ \hat{H}_{\rm rest},
\end{align}
where $\hat{H}_{\rm rest}$ can act on systems other than the clocks as well as contain coupling terms. We can now apply the considerations from Sec.~\ref{ssec:singelclockPW} to each clock individually.
For a given choice of clock, say $C_i$, the solutions to the constraint in Eq.~\eqref{eq:constraint equation} can be written as
\begin{align}
    \label{eq:physical state_general}
    \kket{\Psi}= \frac{1}{2\pi}\int_{\mathds{R}} \, dt_i\ket{t_i}_{C_i} \ket{\psi_{| C_i}(t_i)}_{\lnot C_i} ,
\end{align}
where $\ket{t_i}$ is the eigenstate of a time observable associated with clock $C_i$, as in Eq.~\eqref{eq:Toperator}. The subscript $\lnot C_i$ denotes the Hilbert space containing everything but the clock $C_i$, i.e. $\bigotimes_{i'\neq i} \mathcal{H}_{C_i'}\otimes \mathcal{H}_S$. One can again define the conditional states $\ket{\psi_{|C_i}(\tau_i)}= \langle \tau_i\kket{\Psi}$,
which give the evolution of everything but clock $C_i$, described with respect to $C_i$. Assuming that $\bbra{\Psi}[\proj{t_i}\otimes \mathds{1}_S,\hat{H}_{\rm rest}]\kket{\psi}=0$, this state then evolves unitarily. \\

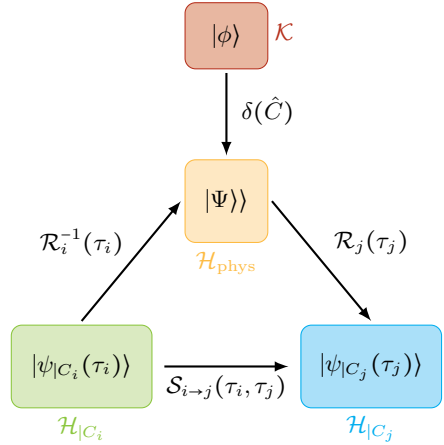
\begin{figure}[h!]
    \centering
\begin{tikzpicture}[scale=0.55]
\draw[draw=BrickRed, fill=BrickRed!30,rounded corners=4pt] (-1,5.2) rectangle (1,6.8);
\node[] (H_phys)at (1.4,6.1){\color{BrickRed}$ \mathcal{K}$};
\node[] (phys)at (0,6){$ \ket{\phi}$};

\draw [-latex,thick] (0,5.1) to (0,3.1);
\node[] (delta)at (1,4.2){$ \delta(\hat{C})$};

\draw[draw=Dandelion, fill=Dandelion!30,rounded corners=4pt] (-1,1) rectangle (1,3);
\node[] (H_phys)at (0,0.5){\color{Dandelion}$ \mathcal{H}_{\rm phys}$};
\node[] (phys)at (0,2){$ \kket{\Psi}$};

\draw[draw=LimeGreen, fill=LimeGreen!30,rounded corners=4pt] (-5.2,-1) rectangle (-1.8,-3);
\node[] (H_phys)at (-3.5,-3.5){\color{LimeGreen}$ \mathcal{H}_{|C_i}$};
\node[] (phys)at (-3.5,-2){$ \ket{\psi_{|C_i}(\tau_i)}$};

\draw [-latex,thick] (-1.1,2)(-3.5,-0.9) to (-1.1,2);
\node[] (Ri)at (-3.5,1){$ \mathcal{R}^{-1}_i(\tau_i)$};

\draw[draw=cyan, fill=cyan!30,rounded corners=4pt] (1.8,-1) rectangle (5.2,-3);
\node[] (H_phys)at (3.5,-3.5){\color{cyan}$ \mathcal{H}_{|C_j}$};
\node[] (phys)at (3.5,-2){$ \ket{\psi_{|C_j}(\tau_j)}$};

\draw [-latex,thick] (1.1,2) to (3.5,-0.9);
\node[] (Rj)at (3.5,1){$ \mathcal{R}_j(\tau_j)$};

\draw[-latex,thick] (-1.5,-2)-- (1.5,-2);
\node[] (map2)at (0,-2.5){$ \mathcal{S}_{i\rightarrow j}(\tau_i ,\tau_j )$};

 \end{tikzpicture}
    \caption{The relationship between the kinematical, physical, and two conditional Hilbert spaces corresponding to the reference frames of clocks $C_i$ and $C_j$. To transform from the reference frame of $C_i$ to that of $C_j$, one first applies the inverse of the reduction map $\mathcal{R}_i(\tau_i)$ which gives the conditional states of $C_i$. This takes us to the Hilbert space of solutions to the constraint equation, namely $\mathcal{H}_{\rm phys}$. One then applies the reduction map $\mathcal{R}_j(\tau_j)$, which is used to obtain the conditional states with respect to clock $C_j$. In total this gives the frame change map $\mathcal{S}_{i\rightarrow j}(\tau_i ,\tau_j )$ according to Eq.~\eqref{eq:framechange}. 
    }
    \label{fig:framechange}
\end{figure}

Scenarios described by the constraint operator in Eq.~\eqref{eq:constraint_general} will give rise to different evolutions when conditioning on different clocks. In other words, the chosen clock system is a temporal reference frame and the induced relational dynamics are frame-dependent. In order to change from one frame to another~\cite{hohn2020switch,vanrietvelde2020change}, we describe the Page-Wootters formalism in terms of \emph{reduction maps}, as in~\cite{Trinity2021}. The map
\begin{align}
    \label{eq:def_R}
    \mathcal{R}_i(\tau_i)= \bra{\tau_i}_{C_i}\otimes\mathds{1}_{\lnot C_i},
\end{align}
takes states from the Hilbert space of the solutions to the constraint equation $\mathcal{H}_{\rm phys}$ to the space of conditional states of clock $C_i$ at time $t_i=\tau_i$, whose Hilbert space we denote~$\mathcal{H}_{|C_i}$. Note that since we are considering only ideal clocks here, the latter space admits a factorization along kinematical lines, i.e.~$\mathcal{H}_{|C_i}\simeq\bigotimes_{i'\neq i} \mathcal{H}_{C_i'}\otimes \mathcal{H}_S$~\cite{ali2022quantum}, as we shall discuss in more detail below. On the other hand, the inverse of the reduction map,
\begin{align}
    \label{eq:def_Rdagger}
    \mathcal{R} ^{-1}_i(\tau_i)=\delta(\hat{C}) \,\ket{\tau_i}_{C_i}\otimes \mathds{1}_{\lnot C_i},
\end{align}
takes a conditional state of clock $C_i$ back to the corresponding solution of the constraint equation. We then combine such maps to change between two temporal reference frames, see Fig.~\ref{fig:framechange}. 
Changing from the frame of clock $i$ to that of clock $j$ is thus achieved via the map
\begin{align}
    \label{eq:framechange}
    \mathcal{S}_{i\rightarrow j}(\tau_i,\tau_j)= \mathcal{R}_j(\tau_j)\mathcal{R}^{-1}_i(\tau_i),
\end{align}
which takes conditional states $\ket{\psi_{| C_i}(\tau_i)}_{\lnot C_i}$ to conditional states $\ket{\psi_{| C_j}(\tau_j)}_{\lnot C_j}= \mathcal{S}_{i\rightarrow j}(\tau_i,\tau_j)\ket{\psi_{| C_i}(\tau_i)}_{\lnot C_i}$. Similarly to the conditional states, an observable $\mathcal{O}_{|C_i}$ relative to $C_i$ transforms under the change of temporal reference frame to the observable $\mathcal{O}_{|C_j}=\mathcal{S}_{i\rightarrow j}\mathcal{O}_{|C_i}\mathcal{S}^{\dagger}_{i\rightarrow j}$ (where we have omitted the $\tau_i$ and $\tau_j$ labels for simplicity). As discussed in~\cite{ali2022quantum}, conditioning on different clocks will induce different subspace structures for the remaining Hilbert space. We illustrate below how this leads observables local to a certain subsystem when described according to one clock, to become non-local (i.e. acting on multiple subsystems) when described according to another clock~\cite{ali2022quantum}.

Consider, for example, two clocks $C_1$ and $C_2$, and a system $S$ comprised of three subsystems, i.e. $\mathcal{H}_S=\mathcal{H}_A\otimes\mathcal{H}_B\otimes\mathcal{H}_C$, and no interactions between the clocks and $S$. In this case we are guaranteed that 
\begin{equation}
\mathcal{H}_{|C_1}\simeq
    \mathcal{H}_{A|C_1}\otimes \mathcal{H}_{B|C_1}\otimes \mathcal{H}_{C|C_1} \simeq \mathcal{H}_{A|C_2}\otimes \mathcal{H}_{B|C_2}\otimes \mathcal{H}_{C|C_2}\simeq\mathcal{H}_{|C_2} ,
\end{equation}
and that the individual subspaces are isomorphic to one another, $\mathcal{H}_{X|C_1}\simeq\mathcal{H}_{X|C_2}$ for $X=A,B,C$ (note that in general, for non-ideal clocks, this does not hold). Importantly, however, despite their isomorphism, $\mathcal{H}_{X|C_1}$ and $\mathcal{H}_{X|C_2}$ are distinct spaces, whose corresponding algebras of observables do not coincide (see Theorem~1 in~\cite{ali2022quantum}).

Nonetheless, if the clocks are ideal, interactions between subsystems do \emph{not} change in the following sense. Consider, for example
\begin{align}
    \label{eq:exmaple_subsystems}
    \hat{C}=\hat{H}_{C_1}+\hat{H}_{C_2}+ \hat{H}_{A}+\hat{H}_{B}+\hat{H}_{C}+\hat{H}_{BC} \, ,
\end{align}
where the system Hamiltonian consists of the free evolution of $A,B$ and $C$ and a coupling term between $B$ and $C$. Noting that for both cases $i=1,2$, the constraint $\hat{C}$ consists of $\hat{H}_{C_i}$ plus some term not acting on $\mathcal{H}_{C_i}$, and one can therefore apply the formalism in Sec.~\ref{ssec:singelclockPW} to condition on either clock $C_i$, finding that the conditional state $\ket{\psi_{| C_i}(t_i)}_{\lnot C_i}$ evolves according to the Hamiltonian 
\begin{equation}
    \label{eq:conditional_Hamiltonian}
    \hat{H}_{\neg C_i|C_i}= \hat{H}_{C_j|C_i}+  \hat{H}_{A|C_i}+\hat{H}_{B|C_i}+\hat{H}_{C|C_i}+\hat{H}_{BC|C_i},
\end{equation}
with $j=1,2$ and $j\neq i$, and where the Hamiltonians $\hat{H}_{X|C_i}$ act only on $\mathcal{H}_{X|C_i}$ and $\hat{H}_{BC|C_i}$ correlates subsystems $B|C_i$ and $C|C_i$. Hence, in either reference frame subsystems $B$ and $C$ are coupled to one another but not to the other systems. However, we stress that $\hat{H}_{\neg C_1|C_1}\neq\hat{H}_{\neg C_2|C_2}$ in general, unless $[\hat{H}_{B}+\hat{H}_{C},\hat{H}_{BC}]=0$, a fact which follows from Corollary~4 of~\cite{Trinity2021}. This may seem counterintuitive, but one should recall that writing $\hat{H}_{\neg C_1|C_1}$ or $\hat{H}_{\neg C_2|C_2}$ in the form of Eq.~\eqref{eq:conditional_Hamiltonian} corresponds to two different factorizations. To summarize, despite the different notion of subsystems in the two frames, the structure of subsystem interactions is frame-independent when the clocks in question are ideal and do not interact with the systems that they are used to describe.

Considering on the other hand an observable $\mathcal{O}_{|C_1}$, which in the frame of $C_1$ acts only on subsystem $C$
\begin{align} \label{eq:factored_nonclock_obs}
    \mathcal{O}_{|C_1}=\mathds{1}_{C_2 AB|C_1}\otimes\hat{O}_{C|C_1}
\end{align}
we find that according to clock $C_2$ this observable acts as
\begin{align}
    \label{eq:obs_transform example1}
    \mathcal{O}_{|C_2}&=\mathcal{S}_{1\rightarrow 2}(\tau_1,\tau_2)\mathcal{O}_{|C_1}\mathcal{S}^{-1}_{1\rightarrow 2}(\tau_1,\tau_2)
    =  \int_{\rm \mathds{R}} ds \proj{\tau_1+s}_{C_1}\otimes \mathds{1}_A \otimes \hat{O}'_{BC}(s),
\end{align}
where $\hat{O}'_{BC}(s)= e^{-is(\hat{H}_{B}+\hat{H}_{C}+\hat{H}_{BC})} (\mathds{1}_{B}\otimes\hat{O}_C) e^{is(\hat{H}_{B}+\hat{H}_{C}+\hat{H}_{BC})}$. Thus, not only does observable $\mathcal{O}_{C_2}$ act non trivially on both subsystems $B$ and $C$ (as a consequence of the coupling term $\hat{H}_{BC}$ in the constraint) but also on clock $C_1$. In particular, Eq.~\eqref{eq:obs_transform example1} shows that an observable which acts independently on some tensor factor according to clock $C_1$, is seen by clock $C_2$ to be correlated with the time observable of clock $C_1$.

\section{A limit to joint temporal localizability and clock synchronization}

\label{sec:clock-synch}

We saw above how, in the multiple-clock scenario, the Page-Wootters formalism leads to reference-frame-dependent relational dynamics. Importantly the relational evolution of one clock with respect to another includes some \emph{necessary time delocalization}, even in the absence of clock-system interactions (cf.~\cite{castro2020quantum}, which examines the time delocalization induced by clock-system interactions). Consider the simplest example of a multiple-clock scenario, namely two clocks $C_1$ and $C_2$ and a third system $S$, i.e.~$\mathcal{K}=\mathcal{H}_{C_1}\otimes\mathcal{H}_{C_2}\otimes\mathcal{H}_S$. When generating a physical state according to Eq.~\eqref{eq:physical_state_from_deltaC}, we can follow a similar procedure to the one used in Sec.~\ref{ssec:singelclockPW}, and write an arbitrary kinematical states $\ket{\phi} \in \mathcal{K}$ in the form 
\begin{align}
  \ket{\phi}= \int_{\mathds{R}} dt_1 \, dt_2 \,\sumint_{\;\lambda} \varphi(t_1,t_2,\lambda)\ket{t_1}_{C_1}\ket{t_2}_{C_2}\ket{\lambda}_S,
  \label{eq:initial state}
\end{align}
where $\varphi(t_1,t_2,\lambda)$ encodes the correlations between the two clocks $C_1$ and $C_2$, as well as with the system $S$, and as before $\lambda$ labels an arbitrary basis of $\mathcal{H}_{S}$. 
If there are no coupling terms between the clocks and the system, i.e. $\hat{C}=\hat{H}_{C_1} +\hat{H}_{C_2} + \hat{H}_S$ the corresponding physical state is
\begin{align}
  \kket{\Psi} &= \frac{1}{2\pi}\int_{\mathds{R}} ds \, dt_1 \, dt_2 \, \sumint_{\;\lambda}\varphi(t_1,t_2 ,\lambda)\ket{t_1+s}_{C_1}\ket{t_2+s}_{C_2}e^{-is \hat{H}_S}\ket{\lambda}_S \, .
  \label{eq:Psi_phys explicit}
\end{align}
Recalling that we can write the inner product on the physical Hilbert space in terms of projectors onto the time eigenstates~\cite{Smith2019,Trinity2021}:
\begin{equation}
    \braket{\langle\Psi}{\Psi\rangle}_\mathrm{phys}=\bra{\langle\Psi}(\ketbra{\tau_{1}}{\tau_{1}}\otimes\mathds{1}_{C_{2}}\otimes\mathds{1}_S)\ket{\Psi\rangle}=\bra{\langle\Psi}(\mathds{1}_{C_{1}}\otimes\ketbra{\tau_{2}}{\tau_{2}}\otimes\mathds{1}_S)\ket{\Psi\rangle} \quad \forall\,\tau_{1},\,\tau_{2},
\end{equation}
we see that normalizability of the physical states implies normalizability of the corresponding conditional states, i.e. that ${\braket{\psi_{|C_i}(\tau_{i})}{\psi_{|C_i}(\tau_{i})} < \infty}$, for all clock times $\tau_i$ and $i\in \{1,2\}$. As we will show now this implies the impossibility of \emph{perfectly synchronized} clocks. By perfect synchronization we mean that there is effectively only one time parameter simultaneously determining the time states for both clocks, i.e.~$\varphi(t_1,t_2,\lambda)= \varphi(t_1,\lambda)\delta(t_1-t_2)$ for some function $\varphi(t_1,\lambda)$, giving a kinematical state of the form $\ket{\phi}= \int_{\mathds{R}} dt \,\sumint_{\lambda} \varphi(t,\lambda)\ket{t}_{C_1}\ket{t}_{C_2}\ket{\lambda}_S$. Applying this condition to Eq.~\eqref{eq:Psi_phys explicit} does not lead to normalizable physical states since their inner product is proportional to the $\delta$-distribution:
 \begin{align}
 \label{eq:perfect_sync_norm}
    \kket{\Psi} &= \frac{1}{2\pi}\int_{\mathds{R}} ds \, dt\,\sumint_{\;\lambda} \varphi(t,\lambda)\ket{t+s}_{C_1}\ket{t+s}_{C_2}e^{-is \hat{H}_S}\ket{\lambda}_S \, , \\
    \langle \langle \Psi \kket{\Psi}_\mathrm{phys}&= \frac{1}{2\pi}\int_{\mathds{R}} ds\, dt\, dt' \, \sumint_{\;\lambda,\lambda'}
    \varphi^*(t',\lambda') \varphi(t,\lambda) \delta^2(t+s-t') \bra{\lambda'}e^{is\hat{H}_S}\ket{\lambda} \propto \delta(0)
    \, . \nonumber
\end{align}
Thus, since $\langle \langle \Psi \kket{\Psi}_\mathrm{phys}$ diverges, the conditional states obtained from $\kket{\Psi}$ are not normalizable at all clock times:~$\braket{\psi_{|C_i}(\tau_i)}{\psi_{|C_i}(\tau_i)}\propto \delta(0)$  $\forall \tau_i,\, i=1,2$. Hence, normalizable physical states --- and consequently conditional states --- require that in the frame of one clock, another clock is necessarily \emph{time-delocalized} to some extent. 

As in Sec.~\ref{ssec:singelclockPW} for the case of a single clock system, we can generate examples of physical states from (non-normalizable) kinematical states of the form $\ket{\phi}= \ket{t_0}_{C_1}\int dt\, \varphi(t,t_0) \ket{t}_{C_2} \ket{\phi_0}_S$. Now $\varphi(t,t_0)$ describes the time delocalization of $C_2$ conditioned on $C_1$ being in state $\ket{t_0}$. We have then ``shifted'' the necessary time delocalization to clock $C_2$ only. To simplify things further, we will choose $t_0=0$, and denote $\varphi(t)\coloneqq\varphi(t,0)$ in what follows. If we then want the clocks to ``read the same time at $t=0$'', the function $\varphi(t)$ should be sharply peaked around $0$. 

Since the divergence in Eq.~\eqref{eq:perfect_sync_norm} arises due to our assumption of ideal clocks, one may ask if the limit to clock synchronization continues to hold outside of this assumption. Consider for a moment non-ideal clocks, such that $\chi(t-t')\coloneqq\braket{t}{t'}\neq\delta (t-t')$ (see~\cite{Trinity2021}, Sec.~III). In the cases where $\chi(0)$ is finite, e.g. continuous-spectrum clocks with a doubly-bounded spectrum or discrete-spectrum clocks with a spectrum bounded below, we can normalize the physical states $\kket{\Psi}$ in Eq.~\eqref{eq:perfect_sync_norm}. However, in each temporal reference frame, the other clock will still appear delocalized in time, i.e. $\ket{\psi_{|C_i}(\tau_i)}=\braket{\tau_i}{\Psi}\rangle$ does not contain a sharp value on clock $C_{j\neq i}$.

\section{Frame-dependent evolution and ordering}
\label{sec:time-reversal}
 We now investigate the extent to which relational dynamics permits disagreement between reference frames about the direction of time and the ordering of operations. We consider a simple example which illustrates the limited extent to which these features are permitted. Consider the above scenario of two ideal clocks and a system $S$, where now the two clocks run in opposite directions, which is equivalent to writing
\begin{align}
    \label{eq:constraint_example_clock_dep_order}
    \hat{C}=\hat{H}_{C_1} - \hat{H}_{C_2} + \hat{H}_S.
\end{align}
As discussed in Sec.~\ref{sec:clock-synch}, we can describe the situation where the two clock are approximately synchronized at $t=0$ by using Eq.~\eqref{eq:physical_state_from_deltaC} and a kinematical state of the form $\ket{\phi}= \ket{t_0=0}_{C_1}\int dt \varphi(t) \ket{t}_{C_2} \ket{\phi_0}_S$, where $\varphi(t)$ is sharply peaked around $t=0$, giving the following physical and conditional states
\begin{align}
\kket{\Psi}&= \frac{1}{2\pi}\int_{\mathds{R}} ds \, dt\, \varphi(t)e^{-is\hat{C}} \ket{0}_{C_1}\ket{t}_{C_2}\ket{\phi_0}_S
=\frac{1}{2\pi}\int_{\mathds{R}} ds \, dt\, \varphi(t) \ket{s}_{C_1}\ket{t-s}_{C_2}e^{-is\hat{H}_S}\ket{\phi_0}_S \, ,
\label{example_state}\\
&\ket{\psi_{| C_1}(\tau_1)}=\langle \tau_1\kket{\Psi}=\int dt \, \varphi(t) \ket{t-\tau_1}_{C_2}e^{-i\tau_1\hat{H}_S} \ket{\phi_0}_S \, ,\\
&\ket{\psi_{| C_2}(\tau_2)}=\langle \tau_2\kket{\Psi}= \int dt \, \varphi(t) \ket{t-\tau_2}_{C_1}e^{i(\tau_2-t)
\hat{H}_S} \ket{\phi_0}_S \, ,
\label{example_perspectives}
\end{align}
whence we can see that the two different clocks observe the system evolving in two different directions in time i.e.~the two frames are defined such that their clocks run in opposing directions. However, the temporal correlations between the system and the remaining clock are different in the two reference frames due to the time delocalization. In particular, despite choosing opposing clocks in this manner, we do not obtain frames which are a perfect time-reversal of each other. To understand this better, let us consider the dynamics of the system alone in each temporal reference frame, i.e. 
\begin{align}
\rho_{S|C_1} (\tau_1) &:= \text{Tr}_{C_2} \left( \ket{\psi_{| C_1}(\tau_1)}\bra{\psi_{| C_1}(\tau_1)} \right)= e^{-i\tau_1 \hat{H}_S} \proj{\phi_0}_{S} e^{i\tau_1 \hat{H}_S}
\label{Perspectival_states1}
 \\
\rho_{S|C_2} (\tau_2) &:= \text{Tr}_{C_1} \left( \ket{\psi_{| C_2}(\tau_2)}\bra{\psi_{| C_2}(\tau_2)} \right) = \int dt\, |\varphi(t)|^2 e^{i(\tau_2-t) \hat{H}_S} \proj{\phi_0}_{S} e^{-i(\tau_2-t) \hat{H}_S} = e^{i\tau_2 \hat{H}_S}\rho_{S|C_2} (0)e^{-i\tau_2 \hat{H}_S}.
\label{Perspectival_states2}
\end{align}
These are the reduced states of the system according to the two clocks, showing the opposing directions of time as well as time delocalization in the frame of clock $C_2$. However, recalling that $\varphi(t)$ is sharply peaked around $t=0$ we obtain  at initial time $\tau_1=\tau_2=0$ 
\begin{align}
\rho_{S|C_1} (0) &= \proj{\phi_0}_{S}
\quad \text{and} \quad \rho_{S|C_2} (0) =  \int dt\, |\varphi(t)|^2  e^{-it \hat{H}_S} \proj{\phi_0}_{S} e^{it \hat{H}_S} = \proj{\phi_0}_{S}+\hat{\varepsilon}, 
\label{eq:initial state example}
\end{align}
where $\hat{\varepsilon}\coloneq\int dt\, |\varphi(t)|^2  e^{-it \hat{H}_S} \proj{\phi_0}_{S} e^{it \hat{H}_S} - \proj{\phi_0}_{S}$ is such that $|\bra{\psi_{S|C_{2}}}\hat{\varepsilon}\ket{\psi_{S|C_{2}}}|<<1$ $\forall\,\ket{\psi_{S|C_{2}}}\in\mathcal{H}_{S|C_{2}}$. While the difference between the states according to the two clocks at $\tau_1=\tau_2=0$ can in principle be arbitrarily small, if the system $S$ exhibits the quantum analogue of the ``butterfly effect'' in classical chaotic systems~\cite{cotler2018out}, then this difference will be exponentially amplified over time. For large enough $S$, this eventually results in an unavoidable distinction between the past of $S$ according to clock $C_1$, and the future of $S$ according to clock $C_2$, despite the fact that $S$ evolves under $e^{-it \hat{H}_S}$ for $C_1$ and under $e^{it \hat{H}_S}$ for $C_2$. \footnote{Given the connection between quantum chaos and the emergence of thermal behavior~\cite{d2016quantum}, we speculate that it is extremely unlikely (strictly speaking, ``atypical'') for two clocks to be capable of disagreeing about the thermodynamic arrow of time. For example, if $\proj{\phi_0}_{S}$ corresponds to a many-body state whose subsystems have recently thermalized~\cite{d2016quantum}, then according to the description relative to clock $C_1$, the time-reversed evolution of $\proj{\phi_0}_{S}$ entails those subsystems going from thermal to athermal, say at some time $-\tau'$. Now, since $S$ begins in the state $\proj{\phi_0}_{S}+\hat{\varepsilon}$ relative to clock $C_2$, and since non-equilibrium states are highly atypical~\cite{reimann2007typicality}, the exponentially divergent paths of $\proj{\phi_0}_{S}$ and $\proj{\phi_0}_{S}+\hat{\varepsilon}$ under the application of the unitary operation $e^{i\tau' \hat{H}_S}$ suggests that it is extremely unlikely that the subsystems of $\rho_{S|C_2} (\tau')$ are out of thermal equilibrium. Further investigation is needed to determine whether this intuition holds. For an in depth discussion of the reference-frame dependence of thermodynamic properties, see~\cite{hoehn2023quantum}.}\\

Outside of the chaotic case discussed above, one can ask if it is possible to contrive an example where the two clocks not only see the same initial state evolve in different directions in time, but in fact see events occurring in opposite orders. One way of defining events quantum mechanically is by the application of quantum operations or Completely Positive trace non-increasing (CP)-maps --- see Sec.~\ref{sec:OperationalCausality&PW} below. Consider two such operations, $\mathcal{E}_A$ and $\mathcal{E}_B$, whose Steinspring dilations can be written as $\mathcal{E}_A[ \cdot ] =\tr_{A^{\prime}} \left(U_A (\cdot\otimes \proj{a_0}_{A^{\prime}}) U_A^{\dagger}\right)$ and $\mathcal{E}_B[ \cdot ]=\tr_{B^{\prime}} \left(U_B (\cdot\otimes \proj{b_0}_{B^{\prime}}) U_B^{\dagger}\right)$. Is it then possible, for the clocks $C_1$ and $C_2$ to assign different orders to these operations, e.g. that $C_1$ sees the evolution as $\mathcal{E}_B \circ\mathcal{E}_A$ acting on some initial state, while $C_2$ sees this as $\mathcal{E}_A \circ\mathcal{E}_B$ acting on the same state? 
As we discuss in Appendix~\ref{app:frame dependent ordering} such a clock-dependent ordering only appears (approximately) in the relational picture described above if the unitary component of the Steinspring dilation of $\mathcal{E}_A$ and $\mathcal{E}_B$ is time-reversal invariant (i.e. Hermitian).

\section{Operational causality and the Page-Wootters formalism}
\label{sec:OperationalCausality&PW}

As stated in Sec.~\ref{sec:introduction}, we will study how the operational notion of causality, which considers causal influences in terms of \emph{quantum operations}~\cite{costa2016quantum,perinotti2021causal,oreshkov2012quantum,chiribella2025maximum}, can be embedded in the Page-Wootters formalism. A quantum operation can causally influence another operation if we can detect the application of the former via the latter. After briefly presenting this operational approach to causality in quantum theory, we show how a first, ad hoc attempt of incorporating it into the Page-Wootters formalism breaks down when considering multiple temporal frames, i.e. clock systems. 

\subsection{The operational approach to causal relations}
\label{ssec:OpertionalCausality}

\emph{Operational causality} identifies causal influences with signaling correlations between quantum operations. A quantum operation is a CP map $\mathcal{M}$ on quantum states, potentially producing a classical outcome. If some hypothetical agent $A$, by applying operation $\mathcal{M}_A$, can affect the probability associated with the outcome of operation $\mathcal{M}_B$, applied by a hypothetical agent $B$, then we say that $A$ can causally influence $B$. 
Whether this is possible will, of course, depend on the underlying dynamics of the quantum systems that these agents act upon, as encoded in the map connecting the output of $A$'s operation to the input of $B$'s operation.

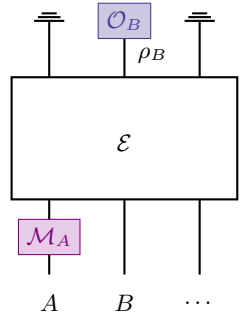
\begin{figure}[h]
    \centering
    \begin{tikzpicture}[scale=0.5]
\draw[thick] (0,-1.25) -- (0,5.5);
\draw[thick] (2,-1.25) -- (2,5.5);
\draw[thick] (4,-1.25) -- (4,5.5);

\node[] (A)at (0,-2){$ A$};
\node[] (B)at (2,-2){$ B$};
\node[] (dots)at (4,-2){$\dots$};

\draw[draw=black,thick,fill=white] (-1,0.75) rectangle (5,4);
\node[] (E)at (2,2.25){$ {\mathcal{E}}$};

\node[draw=violet, fill=violet!20] (intervention)at (0,-0.25){\color{violet}{$\mathcal{M}_A$}};
\node[draw=BlueViolet, fill=BlueViolet!20] (measurement)at (2,5.5){\color{BlueViolet}{$\mathcal{O}_B$}};
\node[] (measuremtn)at (2.75,4.6){$\rho_B$};

\draw[thick] (-0.4,5.5) -- (0.4,5.5);
\draw[thick] (-0.3,5.6) -- (0.3,5.6);
\draw[thick] (-0.2,5.7) -- (0.2,5.7);

\draw[thick] (3.6,5.5) -- (4.4,5.5);
\draw[thick] (3.7,5.6) -- (4.3,5.6);
\draw[thick] (3.8,5.7) -- (4.2,5.7);

 \end{tikzpicture}
    \caption{Operational causality. Let a collection of quantum systems $A,B,\dots$ evolve according to the CPTP map $\mathcal{E}$. We say that the input system $A$ signals to the output system $B$ if there exist a local map $\mathcal{M}_A$ on $A$ and an observable $\mathcal{O}_B$ on $B$ such that the application of $\mathcal{M}_A$ can be detected via $\mathcal{O}_B$. Signaling can occur if the reduced state $\rho_B$ of the output on $B$ differs depending on whether $\mathcal{M}_A$ is applied or not.}
    \label{fig: signalling}
\end{figure}

More concretely, let us consider some dynamical process represented by $\mathcal{E}$, which in general is a completely positive trace preserving (CPTP) map and for convenience (and, in particular, compatibility with the Page-Wootters formalism) can be dilated to a unitary using ancillary systems. We can investigate its causal properties by considering the scenario depicted in Fig.~\ref{fig: signalling}. At an initial time $t_i$ we apply one from a set of quantum operations $\lbrace\mathcal{M}^i_A \rbrace_{i}$ to a sub-system $A$. Later, at a final time  $t_f>t_i$ we apply an operation $\mathcal{M}_B$, which for simplicity we assume to be the measurement of some observable $\mathcal{O}_B$, to a different subsystem $B$, obtaining some outcome $b$.  If we then find that 
\begin{align} \label{eCausalSignalling}
    p(b|\mathcal{M}^{i}_A)\neq p(b|\mathcal{M}^{j}_A),
\end{align}
where $i\neq j$ denote different maps on $A$, then the choice of map by agent $A$ can be detected via the outcomes obtained by agent $B$. Agent $A$ is therefore in the causal past of agent $B$.
To further simplify matters, we will compare the application of an operation $\mathcal{M}_A$ at $t_i$ to the case where no operation is applied, i.e.~$\lbrace\mathcal{M}^i_A \rbrace_{i}=\lbrace\mathds{1}_{A},\mathcal{M}_A\rbrace$. More concretely, given an initial state $\rho(t_i) \in \mathcal{B}( \mathcal{H}_A\otimes \mathcal{H}_B \otimes\dots )$ and the evolution $\mathcal{E}$, when investigating the causal relation between $A$ and $B$ we compare
\begin{align}
    \rho_B (t_f)&=\tr_{\lnot B}\left(\mathcal{E}\left[ \rho(t_i)\right]\right),
\end{align}
where $\tr_{\lnot B}$ denotes the trace over all subsystems except $B$, to 
\begin{align}
    \overline{\rho}_B (t_f)&=\tr_{\lnot B}\left(\mathcal{E}\left[ (\mathcal{M}_A\otimes \mathds{1}_{\lnot A})\rho(t_i)\right]\right).
\end{align}
Thus, $\overline{\rho}_B (t_f)$ and $\rho_B (t_f)$ correspond respectively to the cases where, prior to the evolution $\mathcal{E}$, an intervention on $A$ takes place or not. If these two states of $B$ are \emph{different}, $\overline{\rho}_B (t_f)\neq \rho_B (t_f)$, then there exists an observable $\hat{O}_B= \sum_b b \Pi^b_B$ whose outcome probabilities depend on whether or not the operation $\mathcal{M}_A$ has been applied to $A$ at $t_i$, i.e.
\begin{align}
\label{eq:p(no intervention)}
    &p(b|\text{no intervention})=\tr(\Pi^b_B\rho_B (t_f)) =
    \tr \left( \mathds{1}\otimes \Pi_B^b\mathcal{E}\left[ \rho(t_i)\right]\right)\\
\label{eq:p(intervention)}
    &\neq \tr \left( \mathds{1}\otimes \Pi_B^b\mathcal{E}\left[ \mathcal{M}_A\otimes \mathds{1}\rho(t_i)\right]\right) = \tr(\Pi^b_B\overline{\rho}_B (t_f))=p(b|\text{intervention}), 
\end{align}
meaning that an agent $A$ can signal to another agent $B$ by choosing whether or not to apply operation $\mathcal{M}_A$. A simple illustrative example with two qubits can be found in Appendix~\ref{app:example}.

\subsection{A breakdown of operational causality in relational dynamics}
\label{ssec:OperationalCausalityIssues}

\begin{figure}
    \centering
    \begin{tikzpicture}[scale=0.5]
\node[] (spec)at (-13.5-1,5.75){a)};
\draw[-latex,thick, draw=LimeGreen] (-12.75-1,-2.5)-- (-12.75-1,6.5);
\node[] (C)at (-13.5-1,-2){\color{LimeGreen}$C$};

\draw[thick] (-11-1,-1.25) -- (-11-1,5.5);
\draw[thick] (-9-1,-1.25) -- (-9-1,5.5);
\draw[thick] (-5-1,-1.25) -- (-5-1,5.5);

\node[] (A)at (-11-1,-2){$ A$};
\node[] (B)at (-9-1,-2){$ B$};
\node[] (dots)at (-7-1,-1){$\dots$};
\node[] (X)at (-6,-2){$X$};

\draw[draw=black,thick,fill=white] (-12-1,1) rectangle (-4-1,4);
\node[] (E)at (-8-1,2.5){$ {\mathcal{U}_{|C}}$};

\node[] (dots)at (-8-1,5.5){$\dots$};

\node[] (spec)at (-2.5,5.75){b)};
\draw[-latex,thick, draw=LimeGreen] (-1.75,-2.5)-- (-1.75,6.5);
\node[] (C)at (-2.5,-2){\color{LimeGreen}$C$};

\draw[thick] (0,-1.25) -- (0,5.5);
\draw[thick] (2,-1.25) -- (2,5.5);
\draw[thick] (6,-1.25) -- (6,5.5);

\node[] (A)at (0,-2){$ A$};
\node[] (B)at (2,-2){$ B$};
\node[] (dots)at (4,-1){$\dots$};
\node[] (X)at (6,-2){$X$};

\draw[draw=black,thick,fill=white] (-1,1) rectangle (7,4);
\node[] (E)at (3,2.5){$ {\mathcal{U}_{|C}}$};

\node[] (dots)at (4,5.5){$\dots$};

\node[draw=violet, fill=violet!20] (intervention)at (0,-0.25){\color{violet}{$\mathcal{M}_A$}};
\node[draw=BlueViolet, fill=BlueViolet!20] (measurement)at (2,5.5){\color{BlueViolet}{$\mathcal{O}_B$}};

\draw[thick] (-0.4,5.5) -- (0.4,5.5);
\draw[thick] (-0.3,5.6) -- (0.3,5.6);
\draw[thick] (-0.2,5.7) -- (0.2,5.7);

\draw[thick] (5.6,5.5) -- (6.4,5.5);
\draw[thick] (5.7,5.6) -- (6.3,5.6);
\draw[thick] (5.8,5.7) -- (6.2,5.7);

 \end{tikzpicture}
 \caption{Operational causality within the Page-Wootters formalism: a) The solution to the constraint equation determines the unitary evolution $\mathcal{U}_{|C}$ relative to clock C. By tracing out the ancilla system $X$ we obtain the CPTP map $\mathcal{E}$ we want to study. b) The operation $\mathcal{M}_A$ at some initial time can be absorbed into the choice of kinematical state used to construct a particular solution to the constraint equation $\kket{\Psi^A}=\mathcal{R}^{-1}_C(t_i)\mathcal{M}_A \braket{t_i}{\Psi}\rangle$ giving conditional states  $\ket{\psi^A_{|C}(t)}=\langle t\kket{\Psi^A}=\mathcal{U}_{|C}(t-t_i)\mathcal{M}_A\langle t_i\kket{\Psi}$. The probabilities for the outcomes of observable $\mathcal{O}_B$ at time $t_f$ can then be calculated from the respective conditional state $\ket{\psi^A_{|C}(t_f)}$. Investigating causal properties then, in general, means comparing said probabilities for different operations $\mathcal{M}_A$ and $\mathcal{M}_{A'}$ respectively.
    \label{fig: signallingPW}}
\end{figure}
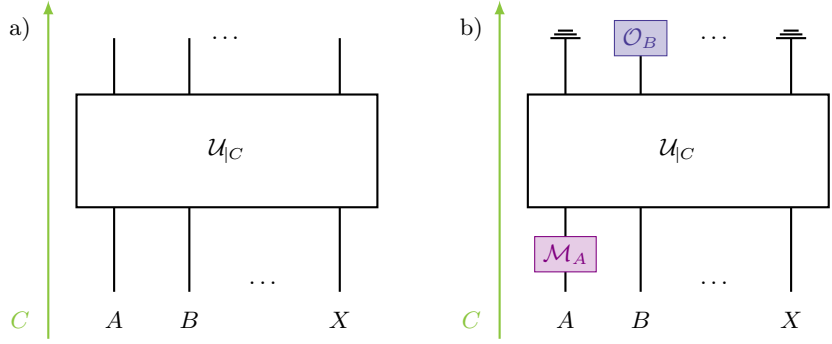

We now describe a naive embedding of operational causality into the Page-Wootters formalism, and show its failure in the multiple-clock setting. First, note that any dynamical evolution, i.e. the CPTP map $\mathcal{E}$, can be represented relationally via a constraint Hamiltonian as in Eq.~\eqref{eq:Schrödingerconstraint}, where an ideal clock $C$ acts as the temporal reference frame, in the following manner. The subsystems $A,B, \dots$, and if necessary an additional ancilla system $X$, are combined into a single kinematical system $\mathcal{H}_{S}\simeq\mathcal{H}_{A}\otimes\mathcal{H}_{B}\otimes\ldots\otimes\mathcal{H}_{X}$, with Hamiltonian $\hat{H}_{S}$, leading to the relational evolution with respect to clock $C$ given by
\begin{equation}
    \label{eq:PW_cond.evolution}
    \mathcal{U}_{|C}(t)\coloneqq e^{-it\hat{H}_S}
\end{equation}
which is such that $\mathcal{U}_{|C}(\Delta t)$ is a unitary corresponding to a Stinespring dilation of the CPTP map $\mathcal{E}$ in Fig.~\ref{fig: signalling}. We can then choose a kinematical state $\ket{t_0}_C\ket{\phi_0}_S$ and, following Sec.~\ref{ssec:singelclockPW}, write an arbitrary solution to the constraint equation as
\begin{align}
    \label{eq:PWsignaling_state}
    &\kket{\Psi} = \frac{1}{2\pi}\int_{\mathds{R}} ds \, \ket{t_0+s}_C e^{-is\hat{H}_{S}} \ket{\phi_{0}}_S \, , \\
    &\qquad \qquad \qquad \text{ with }  \nonumber\\
    &\ket{\psi_{|C} (t)}_S = \langle t \kket{\Psi}=\mathcal{U}_{|C}(t-t_0)\ket{\phi_{0}}_S \, . \label{eq:PWsignaling_red_state}
\end{align}
such that $\tr_{X} (\mathcal{U}_{|C}(\Delta t)\ketbra{\phi_{0}}{\phi_{0}}_{S}\mathcal{U}^{\dagger}_{|C}(\Delta t))=\mathcal{E}[\rho]$, where $\rho$ can be any state on $\mathcal{H}_{A}\otimes\mathcal{H}_{B}\otimes\ldots$,\footnote{Note that the additional system $X$ plays the role of purifying this $\rho$, as well as providing an additional system for the Stinespring dilations of the pertinent CPTP maps.} thus encoding the evolution whose causal properties we wish to study. Now, to actually investigate its causal properties, we have to introduce an intervention. Let us therefore define a conditional state which differs from the one in Eq.~\eqref{eq:PWsignaling_red_state} by the application of an operation $\mathcal{M}_A$ to subsystem $A$ at time $t_i$, namely
\begin{equation}
   \ket{\psi^{A}_{|C} (t)}_S:= \mathcal{U}_{|C}(t-t_i)U_A \braket{t_i}{\Psi}\rangle \, ,
   \label{eq:def_persp_intervention}
\end{equation}
where $U_A$ corresponds to a Stinespring dilation of $\mathcal{M}_A$, whose ancilla system is included in $X$. Note that while we have defined the state in the case of an intervention directly on the reduced space of $S$ relative to the clock (i.e. Eq.~\eqref{eq:def_persp_intervention}), this corresponds to a different choice of physical state $\kket{\Psi^{A}}\coloneqq\mathcal{R}^{-1}_C(t_i)\mathcal{M}_A \braket{t_i}{\Psi}\rangle\neq\kket{\Psi}$, where $\mathcal{R}^{-1}_C(t)$ is the inverse Page-Wootters reduction map defined in Eq.~\eqref{eq:def_Rdagger}.

A causal influence of enacting the operation $\mathcal{M}_A$ can be detected in another subsystem, say $B$, at some final time $t_f=t_i+\Delta t$ according to the clock, if
\begin{equation}
\label{eq:effective_stateB}
    \rho_B^{A}(t_f)=\tr_{\lnot B}\left(\proj{\psi^{A}_{|C}(t_f)}\right) \neq \tr_{\lnot B}\left(\proj{\psi_{|C}(t_f)}\right)=\rho_B(t_f)\, ,
\end{equation}
where, again, $\tr_{\lnot B}$ denotes the trace over all subsystems but $B$. If these two reduced states differ from one another, then there exists an observable $\hat{O}_B= \sum_b b \Pi^b_B$ on subsystem $B$ such that 
\begin{align}
\label{eq:P(intervention1)}
    &p_{|C}(b|\mathcal{M}_A)=\bra{\psi^{A}_{|C}(t_f)} \mathds{1} \otimes \Pi^b_B \ket{\psi^{A}_{|C}(t_f)} 
    \neq\bra{\psi_{|C}(t_f)} \mathds{1} \otimes \Pi^b_B \ket{\psi_{|C}(t_f)}=p_{|C}(b|\mathds{1})\, ,
\end{align}
where $p_{|C}(b|\mathcal{M}_A)$ denotes the probability of outcome $b$ given intervention $\mathcal{M}_A$, evaluated in the reference frame of clock $C$. We thus obtain the no-intervention and intervention conditional probabilities in Eqs.~\eqref{eq:p(no intervention)} and~\eqref{eq:p(intervention)} in a relational setting. An explicit example of incorporating operational causality into the Page-Wootters formalism in the manner described above is given in Appendix~\ref{app:example}. 

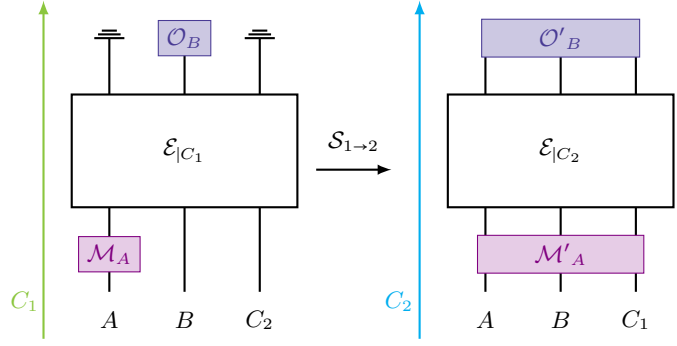
\begin{figure}
    \centering
\begin{tikzpicture}[scale=0.5]


\draw[-latex,thick, draw=LimeGreen] (-1.75,-2.5)-- (-1.75,6.5);
\node[]at(-2.25,-1.5){\color{LimeGreen}$C_1$};
\draw[thick,] (0,-1.25) -- (0,5.5);
\draw[thick,] (2,-1.25) -- (2,5.5);
\draw[thick,] (4,-1.25) -- (4,5.5);
\node[] (A)at (0,-2){$ A$};
\node[] (B)at (2,-2){$ B$};
\node[] (C2)at (4,-2){$C_2$};

\draw[draw=black,thick,fill=white] (-1,1) rectangle (5,4);
\node[] (E)at (2,2.5){$ {\mathcal{E}_{|C_1}}$};

\node[draw=violet, fill=violet!20] (intervention)at (0,-0.25){\color{violet}{$\mathcal{M}_A$}};
\node[draw=BlueViolet, fill=BlueViolet!20] (measuremtn)at (2,5.5){\color{BlueViolet}{$\mathcal{O}_B$}};

\draw[thick] (-0.4,5.5) -- (0.4,5.5);
\draw[thick] (-0.3,5.6) -- (0.3,5.6);
\draw[thick] (-0.2,5.7) -- (0.2,5.7);

\draw[thick] (3.6,5.5) -- (4.4,5.5);
\draw[thick] (3.7,5.6) -- (4.3,5.6);
\draw[thick] (3.8,5.7) -- (4.2,5.7);


\draw[-latex,thick, draw=cyan] (8.25,-2.5)-- (8.25,6.5);
\node[]at(7.7,-1.5){\color{cyan}$C_2$};
\draw[thick,] (10,-1.25) -- (10,5.5);
\draw[thick,] (12,-1.25) -- (12,5.5);
\draw[thick,] (14,-1.25) -- (14,5.5);
\node[] (A)at (10,-2){$ A$};
\node[] (B)at (12,-2){$ B$};
\node[] (C2)at (14,-2){$C_1$};

\draw[draw=black,thick,fill=white] (9,1) rectangle (15,4);
\node[] (E)at (12,2.5){$ {\mathcal{E}_{|C_2}}$};

\draw[-latex,thick] (5.5,2)-- (7.5,2);
\node[] (map2)at (6.5,2.75){$ \mathcal{S}_{1\rightarrow 2}$};
\node[draw=violet, fill=violet!20] (intervention)at (12,-0.25){\color{violet}{$\qquad \mathcal{M'}_A \qquad$}};
\node[draw=BlueViolet, fill=BlueViolet!20] (measurement)at (12,5.5){\color{BlueViolet}{$\qquad \mathcal{O'}_B\qquad$}};

\end{tikzpicture}
    \caption{System-localization of operations for multiple clock systems: If a quantum operation is applied to a single subsystem in one temporal reference frame, here $C_1$, in the reference frame of another clock, $C_2$, this quantum operation acts on multiple subsystems. The change of temporal reference frame $\mathcal{M}_{|C_2}=\mathcal{S}_{1\rightarrow 2}\mathcal{M}_{|C_1}\mathcal{S}^{-1}_{1\rightarrow 2}$ does, in general, not preserve the subsystem localization of operations.}
    \label{fig: PW_intervention}
\end{figure}

This approach, however, encounters a fundamental problem when we
consider multiple clock systems. As we will show below, a Page-Wootters representation well suited for investigating the causal properties according to one clock, is no longer useful to do so after a change of temporal reference frame (i.e. clock system).

Similar to before, we now consider a constraint operator describing two ideal clocks $C_1$ and $C_2$ and a system comprised of multiple subsystems and ancillas $\mathcal{H}_S=\mathcal{H}_A \otimes \mathcal{H}_B \dots \otimes \mathcal{H}_X$
\begin{equation}
    \label{eq:PWsignaling2}
    \hat{C}= \hat{H}_{C_1} +\hat{H}_{C_2} +\hat{H}_{S}\,.
\end{equation}
Analogous to the scenario with a single clock we can construct a physical state as follows:
\begin{align}
    \kket{\Psi}&= \delta(\hat{C})\ket{\phi}= \frac{1}{2\pi}
    \int_{\mathds{R}} ds \, dt \, \varphi(t) \ket{t_0+s}_{C_1} \ket{t+s}_{C_2} e^{-is\hat{H}_{S}} \ket{\phi_0}_S,
\end{align}
where $\varphi(t)$ ensures the necessary time delocalization between the two clocks (see Sec.~\ref{sec:clock-synch}), and $\ket{\phi_0}_S$ is chosen such that we obtain $\tr_{X} (\mathcal{U}_{|C_i}(\Delta t)\ketbra{\phi_{0}}{\phi_{0}}_{S}\mathcal{U}^{\dagger}_{|C_i}(\Delta t))=\mathcal{E}_{|C_i}[\rho]$ for the input state $\rho$ we want to consider. Note, that in general the CPTP map $\mathcal{E}$ will be different for the two reference frames but can be made approximately equal by choosing $\varphi(t)$ sharply peaked around zero (cf.~Sec.~\ref{sec:time-reversal}).
However, if we now try to model quantum operations as in the single-clock scenario, we encounter a problem due to the relativity of subsystems discussed in Sec.~\ref{ssec:multiclocksPW}. To show this it suffices to consider only two subsystems $A$ and $B$ whose causal relationship we wish to investigate, i.e. $\mathcal{H}_S=\mathcal{H}_A \otimes \mathcal{H}_B \otimes \mathcal{H}_X$. Now, consider the case where, in the reference frame of clock $C_1$ we apply an operation $\mathcal{M}_{|C_1}$ corresponding to an intervention on subsystem $\mathcal{H}_{A|C_1}$ at some initial time $\tau^i_1$, and then try to detect its effect on subsystem $\mathcal{H}_{B|C_1}$ by applying an observable $\mathcal{O}_{|C_1}$ at a later time $\tau^f_1$. For these operations to fit into the causal rubric described in Sec.~\ref{ssec:OpertionalCausality}, they have the following form according to $C_1$:
\begin{align}
    \label{eq:MA_OB}
    \mathcal{M}_{|C_1}=\mathds{1}_{C_2}\otimes \mathcal{M}_A\otimes \mathds{1}_B \otimes \mathds{1}_X \quad \text{ and } \quad \mathcal{O}_{|C_1}= \mathds{1}_{C_2}\otimes \mathds{1}_{A} \otimes \mathcal{O}_B  \otimes \mathds{1}_X \, ,
\end{align}
where $\mathcal{M}_A$ and $\mathcal{O}_B$ are to be understood as placeholders for any concrete pair of operation and observable one wants to consider. In the reference frame of clock $C_2$, however, these operations are given by
\begin{align}
    \label{eq:MA_C2}
    \mathcal{M}_{|C_2}&=\mathcal{R}_2(\tau^i_2) \mathcal{R}^{-1}_1 (\tau^i_1)\mathcal{M}_{|C_1}\mathcal{R}_1(\tau^0_1)\mathcal{R}^{-1}_2(\tau^0_2)=\int_{\mathds{R}} ds \; \proj{\tau^0_1+s}_{C_1}\otimes e^{-is\hat{H}_{S}}
    (\mathcal{M}_A\otimes \mathds{1}_B \otimes \mathds{1}_X ) e^{is\hat{H}_{S}} \, , \\
    & \qquad \qquad \qquad\text{ and } \nonumber \\
    \label{eq:OB_C2}
    \mathcal{O}_{|C_2}&=\mathcal{R}_2(\tau^f_2) \mathcal{R}^{-1}_1 (\tau^f_1)\mathcal{O}_{|C_1}\mathcal{R}_1(\tau^f_1)\mathcal{R}^{-1}_2(\tau^f_2)=\int_{\mathds{R}} ds \; \proj{\tau^f_1+s}_{C_1}\otimes e^{-is\hat{H}_{S}}
     (\mathds{1}_A \otimes \hat{O}_B \otimes \mathds{1}_X ) e^{is\hat{H}_{S}} \, ,
\end{align}
where $\tau^i_2$ and $\tau^f_2$ are the initial and final times according to clock $C_2$. Clearly, these operations in general act on all subsystems in the frame of clock $C_2$, namely $\mathcal{H}_{A|C_2}$, $\mathcal{H}_{B|C_2}$ and $\mathcal{H}_{C_1|C_2}$, see Fig.~\ref{fig: PW_intervention} and Appendix\ref{app:example} for a concrete example. 

Since by construction the frame-change map preserves probabilities,  $p_{|C_1}(b|\mathcal{M}_A)=p_{|C_2}(b|\mathcal{M}_A)$, whether or not we observe different probabilities if $\mathcal{M}_A$ is applied or not depends on the physical states $\kket{\Psi}$ and $\kket{\Psi^A}$, and not on the choice of clock. However, observed correlations can be understood in terms of a causal relationship (or lack thereof) between factors $A$ and $B$ with respect to clock $C_1$, in the frame of clock $C_2$, however, no such causal interpretation applies, since there the relevant operations jointly act on all the subsystems.

\section{Interventions in the Page-Wootters formalism}
\label{sec:interventions&PW}

\subsection{Timed interventions via clock-system interaction}

In Sec.~\ref{sec:OperationalCausality&PW}, the presence or absence of an intervention corresponded to alternative choices of physical state (analogous to different initial conditions in a dynamical theory). Here we show how the subsystem-delocalization problem encountered there can be solved by considering a more refined notion of interventions, namely their encoding as dynamical operations, via the constraint operator. Thus one can maintain subsystem-locality across perspectives by explicitly modeling the temporal locality of the interventions. 

To do this, we introduce tailored interaction terms between a clock and the subsystem that a given operation is to be applied to. Let us start by briefly reviewing clock-system interactions within the Page-Wootters formalism~\cite{giovannetti2015quantum,Smith2019,castro2020quantum}. In general such interactions may result in non-unitary evolution of the system with respect to the clock~\cite{paiva2022flow,paiva2022non,hausmann2025measurement,rijavec2025conditions}. However, as shown in~\cite{Smith2019}, clock-system interaction terms of the form 
\begin{align}
    \label{eq:def_H_int}
    \hat{H}_{\rm int}= \int \, dt \, f(t) \proj{t}_C \otimes \hat{K}_S =: \hat{f}(\hat{T})_C \otimes \hat{K}_S
\end{align}
where the time states $\ket{t}$ are defined as in Sec.~\ref{ssec:singelclockPW}, can model unitary relational dynamics with an emergent time-dependent Hamiltonian. As we explain in detail in Appendix~\ref{app:Interaction terms for timed interventions}, using methods from~\cite{castro2020quantum}, constraint operators of the form
\begin{align}
    \label{eq:constraint_with_intervetion}
    \hat{C}= \hat{H}_C +\hat{H}_S+ \hat{f}(\hat{T})_C \otimes \hat{K}_S \;
\end{align}
lead to well defined physical states. Specifically, the elements of the group generated by $\hat{C}$ can be calculated via a Dyson series (cf. Sec.~3.2 in~\cite{Smith2019}), from which one can obtain physical states via Eq.~\eqref{eq:physical_state_from_deltaC}. Here we consider \emph{approximately} instantaneous interventions, in the sense that  $f(t)$ is well-localized around some value $\overline{\tau}$ and can be treated as $\delta(t-\overline{\tau})$ in the Dyson series governing the relational evolution (see Appendix~\ref{app:Interaction terms for timed interventions}). We denote this case by
\begin{align}
    \label{eq:constraint_with_intervetion_delta}
    \hat{C}= \hat{H}_C +\hat{H}_S+ \delta(\hat{T}-\overline{\tau})_C \otimes \hat{K}_S \;
\end{align}
and obtain physical states 
\begin{equation}
    \label{eq:physical_sate_instantanous_intervention}
    \kket{\Psi} =  \frac{1}{2\pi} \int_{-\infty}^{\bar{\tau}} dt \ket{t}_{C} \,e^{-it\hat{H}_{S}}\ket{\phi_0}_{S} + \frac{1}{2\pi} \int_{\bar{\tau}}^{\infty} dt \ket{t}_{C}\,e^{-i(t-\bar{\tau})\hat{H}_{S}}e^{-i\hat{K}_{S}}\,e^{-i\bar{\tau}\hat{H}_{S}}\ket{\phi_0}_{S} ,
\end{equation}
which describe the free evolution of the system according to $\hat{H}_S$ with the application of $e^{-i\hat{K}_{S}}$ at time $\overline{\tau}$. We stress, however, that Eq.~\eqref{eq:constraint_with_intervetion_delta} is an abuse of notation, representing the case of non-zero but negligible delocalization of the function $f(t)$; otherwise, since $\delta^{n}(x)$ is not a well-defined distribution for $n>1$, the exponential of Eq.~\eqref{eq:constraint_with_intervetion_delta} diverges. Eq.~\eqref{eq:physical_sate_instantanous_intervention} should thus be understood as an approximation which neglects the finite width of the function $f(t)$. In~\cite{giovannetti2015quantum} the authors used such instantaneous interventions to describe von Neumann measurements~\cite{vonNeumann1955mathematical,mello2014neumann} within the Page-Wootters formalism, obtaining Eq.~\eqref{eq:physical_sate_instantanous_intervention} in that context (cf.~Eq.~(34) in~\cite{giovannetti2015quantum}). 

To consider general quantum operations, we can again let the system $S$ be comprised of various subsystems $A, B,  \dots X$ and ancillary systems $A', B', \dots $. The subsystems $A, B,  \dots $ represent the factors to which the operations are to be applied, while $A', B' \dots $ provide the necessary additional space for the Stinespring dilations of the relevant CPTP maps ($A'$ for dilating the operation on $A$, and so on). The system $X$, included for generality, allows for a purification of the states which one wishes to embed into the Page-Wootters formalism, as well as a Stinespring dilation of the CPTP map whose causal properties we wish to investigate; for simplicity this will be omitted in the following, as we focus here on the modeling of interventions. Letting $N=A,B,\dots$, the unitary operator dilating a given operation on factor $N$ will be enacted by including the appropriate generator $K_{N N'}$ in the constraint. Specifically the operator $U_{NN'}\coloneqq e^{-i K_{N N^{\prime}}}$ will represent a dilation of the map $\mathcal{M}_N$ on $N$.
In the case where the operation is an observable $\mathcal{O}_{N}$, this corresponds to a von Neumann measurement, correlating $N$ with its ancilla $N^{\prime}$ in the eigenbasis of $\mathcal{O}_{N}$. 

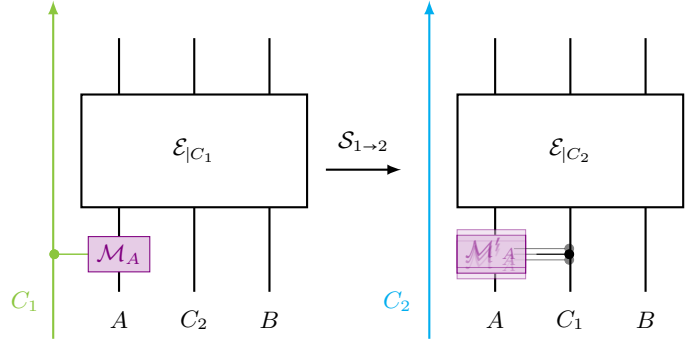
\begin{figure}
    \centering
\begin{tikzpicture}[scale=0.5]

\draw[-latex,thick, draw=LimeGreen] (-1.75,-2.5)-- (-1.75,6.5);
\node[]at(-2.5,-1.5){\color{LimeGreen}$C_1$};
\draw[thick,] (0,-1.25) -- (0,5.5);
\draw[thick,] (2,-1.25) -- (2,5.5);
\draw[thick,] (4,-1.25) -- (4,5.5);
\node[] (A)at (0,-2){$A$};
\node[] (B)at (4,-2){$B$};
\node[] (C2)at (2,-2){$C_2$};

\draw[draw=black,thick,fill=white] (-1,1) rectangle (5,4);
\node[] (U)at (2,2.5){$ {\mathcal{E}_{|C_1}}$};
\draw[Circle-,LimeGreen] (-1.85,-0.25) -- (0,-0.25);

\draw[-latex,thick, draw=cyan] (8.25,-2.5)-- (8.25,6.5);
\node[]at(7.4,-1.5){\color{cyan}$C_2$};
\draw[thick,] (10,-1.25) -- (10,5.5);
\draw[thick,] (12,-1.25) -- (12,5.5);
\draw[thick,] (14,-1.25) -- (14,5.5);
\node[] (A)at (10,-2){$ A$};
\node[] (B)at (14,-2){$ B$};
\node[] (C2)at (12,-2){$C_1$};

\draw[draw=black,thick,fill=white] (9,1) rectangle (15,4);
\node[] (U)at (12,2.5){$ {\mathcal{E}_{|C_2}}$};

\draw[-latex,thick] (5.5,2)-- (7.5,2);
\node[] (map2)at (6.5,2.75){$ \mathcal{S}_{1\rightarrow 2}$};

\draw[Circle-] (12.1,-0.25)--(11.1,-0.25);
\node[draw=violet, fill=violet!20] (intervention)at (9.9,-0.25){\color{violet}{$\mathcal{M'}_A $}};

\draw[Circle-, semitransparent] (12.1,-0.4)--(10.5,-0.4);
\node[draw=violet, fill=violet!20, semitransparent] (intervention)at (9.9,-0.4){\color{violet}{$\mathcal{M'}_A $}};

\draw[Circle-, semitransparent] (12.1,-0.25)--(10.5,-0.25);
\node[draw=violet, fill=violet!20, semitransparent] (intervention)at (9.9,-0.25){\color{violet}{$\mathcal{M'}_A $}};

\draw[Circle-, semitransparent] (12.1,-0.1)--(10.5,-0.1);
\node[draw=violet, fill=violet!20, semitransparent] (intervention)at (9.9,-0.1){\color{violet}{$\mathcal{M'}_A $}};

\draw[Circle-,LimeGreen] (-1.85,-0.25) -- (0,-0.25);
\node[draw=violet, fill=violet!20] (intervention)at (0,-0.25){\color{violet}{$\mathcal{M}_A$}};

 \end{tikzpicture}

    \caption{Timed interventions as part of the constraint operator. The action of a quantum operation on a given subsystems, here $A$, is preserved when changing from the reference frame of clock $C_1$ to the frame of clock $C_2$. However, due to the necessary time delocalization between the two clocks the operations are not be well-localized in time according to the clock which is not coupled to $A$ via the interaction term, here $C_2$.}
    \label{fig:intervention_2clocks}
\end{figure}

We now show how incorporating these operations into the constraint operator preserves subsystem-locality under change of temporal reference frame. Consider again the example of two clocks $C_1$ and $C_2$, two subsystems $A$ and $B$, and the respective ancillas $A'$ and $B'$. An instantaneous intervention on subsystem $A$ at time $\overline{\tau}_1$ according to clock $C_1$ then corresponds to constraint operator
\begin{align}
    \label{eq:constraint-2clocks_1int}
    \hat{C}=\hat{H}_{C_1}+\hat{H}_{C_1} +\hat{H}_{S}+\delta(\hat{T}_1-\overline{\tau}_1)\otimes \hat{K}_{AA'}
\end{align}
where the system Hamiltonian $\hat{H}_S$ act only on $A, B , \dots $. Applying the method described in Appendix~\ref{app:Interaction terms for timed interventions} to Eq.~\eqref{eq:constraint-2clocks_1int}, and using a kinematical state of the form $\ket{\phi}= \int \, dt\, \varphi(t_2) \ket{0}_{C_1}\ket{t_2}_{C_2}\ket{\phi_0}_S$, where $\ket{0}_{C_1}$ denotes $\ket{t_{1}=0}_{C_1}$, gives the following physical and conditional states:
\begin{align}
\label{eq:psi_phys-2clocks_1int}
   \kket{\Psi}=& \frac{1}{2 \pi}\int_{-\infty}^{\overline{\tau}_1} dt_1 \, \int dt_2 \, \varphi(t_2) \ket{t_1}_{C_1}\ket{t_1+t_2}_{C_2} e^{-it_1\hat{H}_{S}}\ket{\phi_0}_S\\
   &+\frac{1}{2 \pi}\int_{\overline{\tau}_1}^{\infty} dt_1 \,  \int dt_2 \, \varphi(t_2) \ket{t_1}_{C_1}\ket{t_1+t_2}_{C_2} e^{-i(t_1-\overline{\tau}_1)\hat{H}_{S}} e^{-i\hat{K}_{AA'}} e^{-i\tau_i\hat{H}_{S}}\ket{\phi_0}_S \nonumber \\
    \ket{\psi_{|C_1}(\tau_1)}=& \braket{\tau_1}{\Psi}\rangle =  \int dt_2 \,\varphi(t_2)\ket{\tau_1+t_2}_{C_2}
        \begin{cases} 
          e^{-i\tau_1\hat{H}_S}\ket{\phi_0}_S & \tau_1< \overline{\tau}_1 \\
           e^{-i(\tau_1-\overline{\tau}_1)\hat{H}_S}U_{AA'} e^{-i\overline{\tau}_1\hat{H}_S}\ket{\phi_0}_S & \overline{\tau}_1 \leq \tau_1 \, ,
       \end{cases}\\
       \ket{\psi_{|C_2}(\tau_2)}=& \braket{\tau_2}{\Psi}\rangle=
        \int dt_2 \,\varphi(t_2)\ket{\tau_2-t_2}_{C_1}
        \begin{cases} 
          e^{-i(\tau_2-t_2)\hat{H}_S}\ket{\phi_0}_S & \tau_2-t_2< \overline{\tau}_1 \\
           e^{-i(\tau_2-t_2-\overline{\tau}_1)\hat{H}_S}U_{AA'} e^{-i\overline{\tau}_1\hat{H}_S}\ket{\phi_0}_S & \overline{\tau}_1 \leq \tau_2-t_2 \, ,
       \end{cases}
\end{align}
Note that now $\mathcal{M}_A$ acts only on subsystem $A$ (via its dilation $U_{AA'}$)
in both temporal reference frames. However, in the frame of $C_2$ whether or not the operation has been applied depends not only on clock time $\tau_2$ but also on the integration variable $t_2$. Hence, the application of $\mathcal{M}_A$ appears instantaneous in the frame of $C_1$, i.e. the clock that times the intervention, but time-delocalized in the reference frame the other clock $C_2$, see Fig.~\ref{fig:intervention_2clocks}. The extent of time-localization depends upon the extent to which $\varphi(t)$ is peaked around a given value, and indeed replacing $\varphi(t)$ with $\delta(t)$ in the above expressions, would give an intervention which is fully system-local and time-local from both perspectives. As discussed in Sec.~\ref{sec:clock-synch}, however, this would lead to divergencies. Recalling that Eq.~\eqref{eq:psi_phys-2clocks_1int} was obtained for a specific form of kinematical state $\ket{\phi}$, we note that in general there is no need for the two perspectives to share some sense of timed intervention, particularly when the kinematical state (or more precisely, the physical state obtained therefrom) does not lead to any correlation between the clock times in the two perspectives. 

In what follows we will present how to use the type of interventions above to consistently incorporate operational causality into the Page-Wootters formalism. Finally, we also show how incorporating timed interventions into the constraint operator allows for describing scenarios with indefinite causal order.

\subsection{Timed interventions and operational causality}
\label{ssec:timedOperations&clock-systemInteractions}

Incorporating the interventions necessary for investigating causality (see Sec.~\ref{ssec:OpertionalCausality}), into the constraint operator, the general form for arbitrarily many clocks and interventions is
\begin{align}
\label{eq:constrain+int}
    \hat{C}= \sum_j \hat{H}_{C_j} +\hat{H}_S + \sum_{jl} \hat{f}(\hat{T}_j)\otimes K_{A_lA^{\prime}_l} ,
\end{align}
where $j$ labels the different clocks, and thus $\hat{T}_j$ refers to the time operator of clock $C_j$. The different subsystems involved in the evolution are labeled $l$ and $l'$ denotes the corresponding ancilla systems, such that $K_{A_l A^{\prime}_l}$ generates the Stinespring dilations of the map to be applied to system $l$. While the approach to investigating causal relations described in Sec.~\ref{ssec:OperationalCausalityIssues} entailed a comparison of different solutions to a single constraint equation, here one compares \emph{different constraints}. These constraints differ only in the operators $K_{A_lA'_l}$ corresponding to different operations $\mathcal{M}_{A_l}$ being applied. From this perspective, investigating causal relations within relational dynamics amounts to comparing representatives of a fixed class of constraint equations and their solutions.

Consider the simplest example, where $S$ consists of two subsystems $A$ and $B$ and their respective ancillas $A'$ and $B'$, and with two approximately instantaneous interventions, one applied to subsystem $A$ followed by the other to subsystem $B$, as timed by a single clock $C$. This corresponds to a constraint operator of the form
\begin{align}
    \label{eq:constriant+int_1clock}
    \hat{C}= \hat{H}_C+\hat{H}_{S}+ \delta(\hat{T}-\tau_i)K_{AA'} +\delta(\hat{T}-\tau_f)K_{BB'},
\end{align}
where $\tau_i<\tau_f$ are the initial and final time according to clock $C$. As before, the Dirac delta notation in Eq.~\eqref{eq:constriant+int_1clock} represents the case where the interventions have some finite but negligible duration centered on $\tau_i$ and $\tau_f$. The class of constraint operators above describes a scenario where $\mathcal{M}_A$ is applied to $A$ at clock-time $\tau_i$ (via its dilation $U_{AA'}\coloneqq e^{-iK_{AA'}}$ on $A$ and $A'$) and $\mathcal{M}_B$ at final clock-time $\tau_f>\tau_i$ (via its dilation $U_{BB'}\coloneqq e^{-iK_{BB'}}$ on $B$ and $B'$). Note that, while the evolution due to $\hat{H}_{S}$ may allow for a bidirectional causal relationship between $A$ and $B$, constraint operators of the form of Eq.\eqref{eq:constriant+int_1clock} correspond to the question of whether $A$ can causally influence $B$ but do not tell us anything about the converse --- to interrogate whether $B$ can influence $A$, one must invert the timings of the interventions.

We will again consider the case where  $\mathcal{M}_B=\mathcal{O}_B$, and compare the situation where some arbitrary $\mathcal{M}_A$ is applied to $A$ with the situation where no intervention on $A$ occurs. The latter case can be obtained by setting $K_{AA'}=0$ in Eq.~\eqref{eq:constriant+int_1clock}, explicitly:
\begin{equation}
    \label{eq:constriant+noint_1clock}
    \hat{C}'= \hat{H}_C+\hat{H}_{S} +\delta(\hat{T}-\tau_f)K_{BB'}.
\end{equation}
Beginning with the case of an arbitrary intervention on $A$, and recalling that in the case of a single clock, an arbitrary physical state can be obtained via a kinematical state of the form $\ket{t=0}_{C}\ket{\phi_0}_{S}$ (see Sec.~\ref{ssec:singelclockPW}), the method in Appendix~\ref{app:Interaction terms for timed interventions} can again be applied to find the physical states generated by Eq.~\eqref{eq:constriant+int_1clock}: 
\begin{align}
    \kket{\Psi^{A}}&= \frac{1}{2 \pi}\left(\int_{-\infty}^{\tau_i} dt\, \ket{t}_C e^{-it\hat{H}_{S}}+ \int_{\tau_i}^{\tau_f} dt\, \ket{t}_C e^{-i(t-\tau_i)\hat{H}_{S}}U_{AA'}e^{-i\tau_i\hat{H}_{S}} \right. \nonumber \\
    &\left.+\int_{\tau_f}^{\infty} dt\, \ket{t}_C e^{-i(t-\tau_f)\hat{H}_{S}} U_{BB'}e^{-i(\tau_f-\tau_i)\hat{H}_{S}}U_{AA'}e^{-i\tau_i\hat{H}_{S}}\right)\ket{\phi_0}_S,
    \label{eq:psi_phys-2int1C}
\end{align}
where the superscript $A$ in $\kket{\Psi^{A}}$ again denotes that an intervention is made on $A$, and we recall that $U_{AA'}$ and $U_{BB'}$ are the unitary dilations of $\mathcal{M}_A$ and $\mathcal{O}_B$ respectively. 
We thus obtain the following expression for the conditional system states
\begin{align}
\label{eq:int_1clock_state}
    &\ket{\psi^{A}_{|C}(\tau)}=\langle \tau \kket{\Psi}=\mathcal{U}^{A}_{|C}(\tau)\ket{\phi_0}, \\
    &\text{ with } \nonumber \\
    \label{eq:int_1clock_unitary}
    &\mathcal{U}^{A}_{|C}(\tau)=
    \begin{cases} 
      e^{-i\tau\hat{H}_{S}} & \tau\leq \tau_i \\
      e^{-i(\tau-\tau_i)\hat{H}_{S}}U_{AA'}e^{-i\tau_i\hat{H}_{S}} & \tau_i\leq \tau\leq \tau_f \\
      e^{-i(\tau-\tau_f)\hat{H}_{S}} U_{BB'}e^{-i(\tau_f-\tau_i)\hat{H}_{S}}U_{AA'}e^{-i\tau_i\hat{H}_{S}} & \tau_f\leq \tau.
   \end{cases}
\end{align}
Note that the timed operations are now part of the overall evolution of the system, and when relating this scenario to the one discussed in Sec.~\ref{ssec:OpertionalCausality} we identify only the evolution in between these operations, i.e. $e^{-i(\tau_f-\tau_i)\hat{H}_{S}}$ in this case, with the CPTP map $\mathcal{E}$ in Fig.~\ref{fig: signalling}.\footnote{Recall that we have omitted the ancilla system $X$, and the CPTP map $\mathcal{E}$ is therefore implicitly unitary in this example. The inclusion of such an ancilla system however does not change any of the following analysis.} As in Eq.~\eqref{eq:P(intervention1)}, we write $p_{|C}(b|\mathcal{M}_A)$ to denote the probability of outcome $b$ given intervention $\mathcal{M}_A$, evaluated in the reference frame of clock $C$, where now the outcome is determined by the von Neumann measurement associated with $\mathcal{O}_B$, performed at $\tau_f$ according to clock $C$. At any time $\tilde{\tau} > \tau_f$, this is
\begin{align*}
    p_{|C}(b|\mathcal{M}_A)&= \bra{\psi^{A}_{|C}(\tilde{\tau})} (\Pi^b_{B'}\otimes \mathds{1}_{\lnot B'}) \ket{\psi^{A}_{|C}(\tilde{\tau})}=  \bra{\phi_0} e^{i\tau_i\hat{H}_{S}} U^{\dagger}_{AA'} e^{i(\tau_f-\tau_i)\hat{H}_{S}} (\Pi^b_B\otimes \mathds{1}_{\lnot B})e^{-i(\tau_f-\tau_i)\hat{H}_{S}}U_{AA'}e^{-i\tau_i\hat{H}_{S}}\ket{\phi_0}\\
    &=\tr\big( (\mathds{1}_A \otimes \Pi^b_B )\mathcal{E} \left[ \mathcal{M}_A \otimes \mathds{1}_B \left[ \rho(\tau_i ) \right]  \right] \big)
    \equiv p(b|\text{intervention}),
\end{align*}
where we have used the fact that, due to the von Neumann measurement, the projection on $B'$ at $\tilde{\tau}>\tau_f$ is equivalent to the corresponding projection onto $B$ at $\tau_f$, and we have identified the input to the map $\mathcal{E}$ as ${\rho(\tau)\coloneqq \tr_{A'B'}(e^{-i\tau\hat{H}_{S}}\proj{\phi_0}e^{i\tau\hat{H}_{S}})}$, recalling again that the more general case of a mixed $\rho(\tau)$ can be handled by the inclusion of an ancilla system $X$. Considering now the case of no intervention, we can replace $U_{AA'}$ with $\mathds{1}_{AA'}$ in $ \kket{\Psi^{A}}$, $\ket{\psi^{A}_{|C}(\tau)}$ and $\mathcal{U}^{A}_{|C}(\tau)$ above, to obtain the no-intervention versions $ \kket{\Psi}$, $\ket{\psi_{|C}(\tau)}$ and $\mathcal{U}_{|C}(\tau)$ respectively, which gives 
\begin{align*}
    p_{|C}(b|\mathds{1}_{A})&= \bra{\psi_{|C}(\tilde{\tau})} (\Pi^b_{B'}\otimes \mathds{1}_{\lnot B'}) \ket{\psi_{|C}(\tilde{\tau})}=  \bra{\phi_0} e^{i\tau_i\hat{H}_{S}}  e^{i(\tau_f-\tau_i)\hat{H}_{S}} (\Pi^b_B\otimes \mathds{1}_{\lnot B})e^{-i(\tau_f-\tau_i)\hat{H}_{S}}e^{-i\tau_i\hat{H}_{S}}\ket{\phi_0}\\
    &=\tr\big( (\mathds{1}_A \otimes \Pi^b_B )\mathcal{E} \left[ \rho(\tau_i ) \right] \big)
    \equiv p(b|\text{no intervention}) .
\end{align*}
We have therefore successfully incorporated the operational notion of causality into the Page-Wootters formalism, with $p_{|C}(b|\mathds{1}_{A})$ and $p_{|C}(b|\mathcal{M}_A)$ interpretable in the same manner as Eq.~\eqref{eq:p(no intervention)} and Eqs.~\eqref{eq:p(intervention)} respectively. However, as we will see below, in contrast with the naive approach that we examined in Sec.~\ref{ssec:OperationalCausalityIssues} (see Eqs.~\eqref{eq:P(intervention1)} in particular), modeling the interventions as part of the constraint operator preserves their suitability for investigating causal relations under the change of temporal frame.

\subsection{Relational causality with multiple clocks}
\label{ssec:RelationalCausality &multiple Clocks}

We now show how, when multiple clocks are considered, the conditional states relative to different clocks can retain an interpretation of a system-local intervention having occurred, and thus infer causal properties from those interventions. Specifically, we consider two clock systems $C_1$ and $C_2$, and distinguish the case where the two interventions of interest are timed by the same clock from those where each clock times one interventions. The latter has the clear advantage that the clock cannot carry a causal influence from $A$ to $B$, and also permits examples with an indefinite ordering of quantum operations, as we will show in Sec.~\ref{sec:RelIndefOrder}.

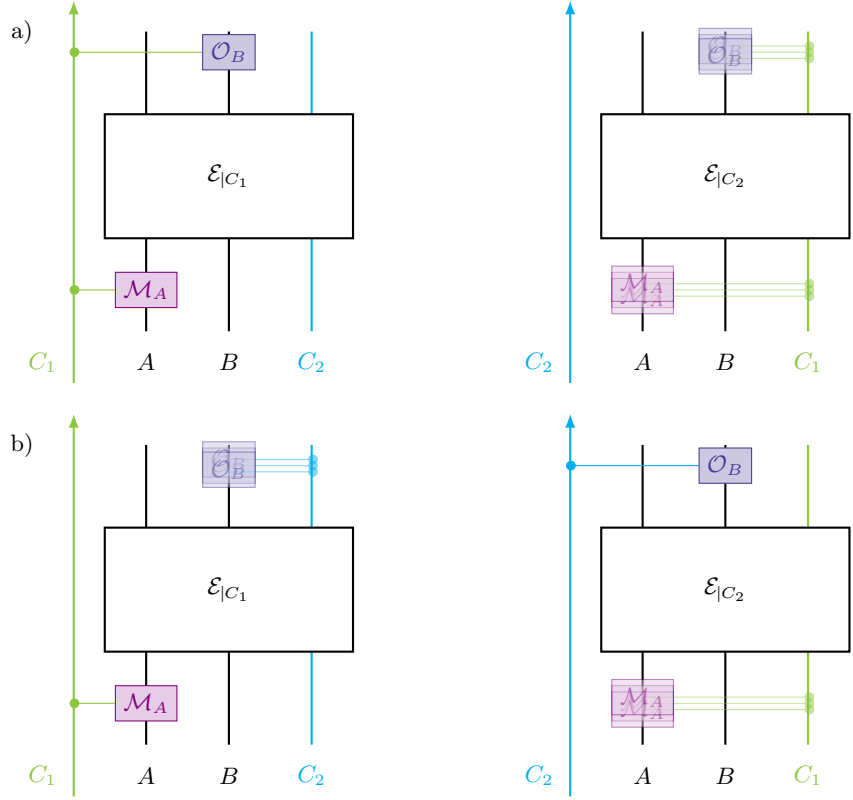
\begin{figure}
    \centering
\begin{tikzpicture}[scale=0.55]

\node[] (a)at (-3,6){a)};
\draw[-latex,thick, draw=LimeGreen] (-1.75,-2.5)-- (-1.75,6.75);
\node[] (C)at (-2.5,-2){\color{LimeGreen}$C_1$};

\draw[thick] (0,-1.25) -- (0,6);
\draw[thick] (2,-1.25) -- (2,6);
\draw[thick, color=cyan] (4,-1.25) -- (4,6);

\node[] (A)at (0,-2){$ A$};
\node[] (B)at (2,-2){$ B$};
\node[] (C2)at (4,-2){\color{cyan}$ C_2$};
\draw[Circle-,LimeGreen] (-1.85,-0.25) -- (0,-0.25);
\draw[Circle-,LimeGreen] (-1.85,5.5) -- (2,5.5);

\draw[draw=black,thick,fill=white] (-1,1) rectangle (5,4);
\node[] (E)at (2,2.5){$ {\mathcal{E}_{|C_1}}$};

\node[draw=violet, fill=violet!20] (intervention)at (0,-0.25){\color{violet}{$\mathcal{M}_A$}};
\node[draw=BlueViolet, fill=BlueViolet!20] (measurement)at (2,5.5){\color{BlueViolet}{$\mathcal{O}_B$}};


\draw[-latex,thick, draw=cyan] (10.25,-2.5)-- (10.25,6.75);
\node[] (C)at (9.5,-2){\color{cyan}$C_2$};

\draw[thick] (12,-1.25) -- (12,6);
\draw[thick] (14,-1.25) -- (14,6);
\draw[thick,LimeGreen] (16,-1.25) -- (16,6);

\node[] (A)at (12,-2){$ A$};
\node[] (B)at (14,-2){$ B$};
\node[] (dots)at (16,-2){\color{LimeGreen}$C_1$};
\draw[Circle-,LimeGreen,semitransparent] (16.15,-0.25) -- (12,-0.25);
\draw[Circle-,LimeGreen,semitransparent] (16.15,5.5) -- (14,5.5);

\draw[draw=black,thick,fill=white] (11,1) rectangle (17,4);
\node[] (E)at (14,2.5){$ {\mathcal{E}_{|C_2}}$};

\node[draw=violet, fill=violet!20,semitransparent] (intervention)at (12,-0.25){\color{violet}{$\mathcal{M}_A$}};

\draw[Circle-, LimeGreen, semitransparent] (16.15,-0.4) -- (12,-0.4);
\node[draw=violet, fill=violet!20, semitransparent] (intervention)at (12,-0.4){\color{violet}{$\mathcal{M}_A $}};

\draw[Circle-, LimeGreen ,semitransparent] (16.15,-0.1) -- (12,-0.1);
\node[draw=violet, fill=violet!20, semitransparent] (intervention)at (12,-0.1){\color{violet}{$\mathcal{M}_A $}};

\node[draw=BlueViolet, fill=BlueViolet!20,semitransparent] (measurement)at (14,5.5){\color{BlueViolet}{$\mathcal{O}_B$}};

\draw[Circle-,LimeGreen, semitransparent] (16.15,5.65) -- (14,5.65);
\node[draw=BlueViolet, fill=BlueViolet!20, semitransparent] (measurement)at (14,5.65){\color{BlueViolet}{$\mathcal{O}_B$}};

\draw[Circle-,LimeGreen, semitransparent] (16.15,5.35) -- (14,5.35);
\node[draw=BlueViolet, fill=BlueViolet!20, semitransparent] (measurement)at (14,5.4){\color{BlueViolet}{$\mathcal{O}_B$}};

\node[] (b)at (-3,6-10){b)};
\draw[-latex,thick, draw=LimeGreen] (-1.75,-2.5-10)-- (-1.75,6.75-10);
\node[] (C)at (-2.5,-2-10){\color{LimeGreen}$C_1$};

\draw[thick] (0,-1.25-10) -- (0,6-10);
\draw[thick] (2,-1.25-10) -- (2,6-10);
\draw[thick,cyan] (4,-1.25-10) -- (4,6-10);

\node[] (A)at (0,-2-10){$ A$};
\node[] (B)at (2,-2-10){$ B$};
\node[] (C2)at (4,-2-10){\color{cyan}$C_2$};
\draw[Circle-,LimeGreen] (-1.85,-0.25-10) -- (0,-0.25-10);
\draw[Circle-,cyan, semitransparent] (4.15,5.5-10) -- (2,5.5-10);

\draw[draw=black,thick,fill=white] (-1,1-10) rectangle (5,4-10);
\node[] (E)at (2,2.5-10){$ {\mathcal{E}_{|C_1}}$};

\node[draw=violet, fill=violet!20] (intervention)at (0,-0.25-10){\color{violet}{$\mathcal{M}_A$}};

\node[draw=BlueViolet, fill=BlueViolet!20,semitransparent] (measurement)at (2,5.5-10){\color{BlueViolet}{$\mathcal{O}_B$}};

\draw[Circle-,cyan, semitransparent] (4.15,5.65-10) -- (2,5.65-10);
\node[draw=BlueViolet, fill=BlueViolet!20, semitransparent] (measurement)at (2,5.65-10){\color{BlueViolet}{$\mathcal{O}_B$}};

\draw[Circle-, cyan, semitransparent] (4.15,5.35-10) -- (2,5.35-10);
\node[draw=BlueViolet, fill=BlueViolet!20, semitransparent] (measurement)at (2,5.4-10){\color{BlueViolet}{$\mathcal{O}_B$}};

\draw[-latex,thick, draw=cyan] (10.25,-2.5-10)-- (10.25,6.75-10);
\node[] (C)at (9.5,-2-10){\color{cyan}$C_2$};

\draw[thick] (12,-1.25-10) -- (12,6-10);
\draw[thick] (14,-1.25-10) -- (14,6-10);
\draw[thick,LimeGreen] (16,-1.25-10) -- (16,6-10);

\node[] (A)at (12,-2-10){$ A$};
\node[] (B)at (14,-2-10){$ B$};
\node[] (dots)at (16,-2-10){\color{LimeGreen}$C_1$};
\draw[Circle-,LimeGreen,semitransparent] (16.15,-0.25-10) -- (12,-0.25-10);
\draw[Circle-,cyan] (10.15,5.5-10) -- (14,5.5-10);

\draw[draw=black,thick,fill=white] (11,1-10) rectangle (17,4-10);
\node[] (E)at (14,2.5-10){$ {\mathcal{E}_{|C_2}}$};

\node[draw=violet, fill=violet!20,semitransparent] (intervention)at (12,-0.25-10){\color{violet}{$\mathcal{M}_A$}};

\draw[Circle-, LimeGreen, semitransparent] (16.15,-0.4-10) -- (12,-0.4-10);
\node[draw=violet, fill=violet!20, semitransparent] (intervention)at (12,-0.4-10){\color{violet}{$\mathcal{M}_A $}};

\draw[Circle-, LimeGreen ,semitransparent] (16.15,-0.1-10) -- (12,-0.1-10);
\node[draw=violet, fill=violet!20, semitransparent] (intervention)at (12,-0.1-10){\color{violet}{$\mathcal{M}_A $}};

\node[draw=BlueViolet, fill=BlueViolet!20] (measurement)at (14,5.5-10){\color{BlueViolet}{$\mathcal{O}_B$}};
 \end{tikzpicture}

    \caption{Operational causality within the Page-Wootters formalism. If the interventions are incorporated into the constraint operator, which subsystem the timed operations are applied to is independent of the temporal reference frame. a) If both interventions are timed by the same clock, $C_1$, they appear well localized in time in the frame of that clock. In the reference frame of $C_2$, however, the operations are time-delocalized and in order investigate causal relations their times of applications have to be separated enough to prevent a temporal overlap. b) If each clock times one of the two operations, time delocalization appears in both temporal reference frame. According to each clock the operation timed by it is well localize in time (by construction) while the operation timed by the other clock appears delocalized in time. }
    \label{fig:causlaity_2clocks}
\end{figure}

The case where both interventions are timed by clock $C_1$ in the manner discussed in Sec.~\ref{ssec:timedOperations&clock-systemInteractions} above, gives constraint operators of the form
\begin{align}
    \hat{C}= \hat{H}_{C_1}+\hat{H}_{C_2}+\hat{H}_{S}+ \delta(\hat{T_1}-\tau_i)K_{AA'} +\delta(\hat{T_1}-\tau_f)K_{BB'}\, ,
\end{align}
which corresponds to the application of some $\mathcal{M}_A$ on $A$ at $\tau_i$ and that of some $\mathcal{M}_B$ on $B$ at $\tau_f$, where both $\tau_i$ and $\tau_f$ are readings of $C_1$. Compared to Eq.~\eqref{eq:constriant+int_1clock}, we have simply added a clock system $C_2$ that evolves independently from everything else. Again applying the method described in Appendix~\ref{app:Interaction terms for timed interventions}, and using a kinematical state of the form $\ket{\phi}= \int \, dt\, \varphi(t_2) \ket{0}_{C_1}\ket{t_2}_{C_2}\ket{\phi_0}_S$ we straightforwardly obtain the following physical states
\begin{align}
\label{eq:Psi_phys_intervention}
   \kket{\Psi}&= \frac{1}{2\pi} \int_{-\infty}^{\tau_i} dt_1 \, \int dt_2 \, \varphi(t_2) \ket{t_1}_{C_1}\ket{t_1+t_2}_{C_2} e^{-it_1\hat{H}_{S}}\ket{\phi_0}_S\\
   &+\frac{1}{2\pi} \int_{\tau_i}^{\tau_f} dt_1 \,  \int dt_2 \, \varphi(t_2) \ket{t_1}_{C_1}\ket{t_1+t_2}_{C_2} e^{-i(t_1-\tau_i)\hat{H}_{S}}U_{AA'}e^{-i\tau_i\hat{H}_{S}}\ket{\phi_0}_S \nonumber\\
    &+\frac{1}{2\pi} \int_{\tau_f}^{\infty} dt_1 \, \int dt_2 \, \varphi(t_2) \ket{t_1}_{C_1}\ket{t_1+t_2}_{C_2} e^{-i(t_1-\tau_f)\hat{H}_{S}} U_{BB'}e^{-i(\tau_f-\tau_i)\hat{H}_{S}}U_{AA'}e^{-i\tau_i\hat{H}_{S}}\ket{\phi_0}_S,\nonumber
\end{align}
where again $\varphi(t_2)$ ensures the necessary time delocalization of one clock with respect to the other (see Sec.~\ref{ssec:multiclocksPW}). Here we omit the superscript $A$ (in contrast with Sec.~\ref{ssec:OperationalCausalityIssues} and Sec.~\ref{ssec:timedOperations&clock-systemInteractions}), as we will compare given interventions in different reference frames, rather than comparing the cases where an intervention does and does not occur. Now, conditioning on clock $C_1$ gives
\begin{align}
    \ket{\psi_{|C_1}(\tau_1)}&=  \int dt_2 \,\varphi(t_2)\ket{\tau_1+t_2}_{C_2} \mathcal{U}_{|C_1}(\tau_1)\ket{\phi_0}_S \, , \\
    &\text{ and } \nonumber \\
     p_{|C_1}(b|\mathcal{M}_A)&= \bra{\psi_{|C_1}(\tilde{\tau})} \Pi^b_{B'}\otimes \mathds{1} \ket{\psi_{|C_1}(\tilde{\tau})}\equiv  p(b|\text{intervention}) \, ,
\end{align}
where $\mathcal{U}_{|C_1}(\tau_1)$ is defined as in Eq.~\eqref{eq:int_1clock_unitary}, and $\tilde{\tau}>\tau_f$. Similarly, when conditioning on $C_2$ we find
\begin{align}
    \ket{\psi_{|C_2}(\tau_2)} &=  \int dt_2 \, \varphi(t_2)\ket{\tau_2-t_2}_{C_1} \mathcal{U}_{|C_2}(\tau_2-t_2)\ket{\phi_0}_S \, , \label{eq:CondStateDelocC2} \\
    &\text{ and } \nonumber \\
    \label{eq:P^2_2clock_intervention_C1}
    p_{|C_2}(b|\mathcal{M}_A)&= \bra{\psi_{|C_2}(\tilde{\tau})} \Pi^b_{B'}\otimes \mathds{1} \ket{\psi_{|C_2}(\tilde{\tau})}\\
    &= \int dt_2 \, |\varphi (t_2)|^2 \bra{\phi_0}\mathcal{U}_{|C_2}(\tilde{\tau}-t_2)^{\dagger} \Pi^b_{B'}\otimes \mathds{1} \mathcal{U}_{|C_2}(\tilde{\tau}-t_2)\ket{\phi_0}_S, \nonumber
\end{align}
where $\mathcal{U}_{|C_2}(\tau)$ is again given by Eq.~\eqref{eq:int_1clock_unitary}. The evolution of the system and thus the probabilities now depend on the time delocalization between the two clocks. In a similar manner to the example considered in Sec.~\ref{sec:interventions&PW}, the probability in Eq.~\eqref{eq:P^2_2clock_intervention_C1} cannot in general be interpreted as arising from a sequence of events occurring at well defined times according to $C_2$. In particular, $\ket{\psi_{|C_2}(\tau_2>\tau_f)}$ contains a contribution where $U_{BB'}$ has been applied (for $\tau_2- t_2>\tau_f$) and another contribution where it has not (for $\tau_2-t_2<\tau_f$). However, assuming that clocks $C_1$ and $C_2$ are jointly localized such that $\varphi(t_2)$ is negligible outside of a region of width $\tilde{\tau}-\tau_f$, then Eq.~\eqref{eq:P^2_2clock_intervention_C1} can be interpreted as an intervention on $A$, an evolution, and then a measurement on $B$, with the caveat that these are subject to some time delocalization, as illustrated in Fig.~\ref{fig:causlaity_2clocks}\hyperref[fig:causlaity_2clocks]{(a)}. \\

Let us now consider a scenario in which the two operations are \emph{timed by different clocks} corresponding to constraint operators
\begin{align}
     \label{eq:2clock_intervention}
    \hat{C}= \hat{H}_{C_1}+\hat{H}_{C_2}+\hat{H}_{S}+ \delta(\hat{T_1}-\tau_a)K_{AA'} +\delta(\hat{T_2}-\tau_b)K_{BB'},
\end{align}
where $\tau_a,\tau_b>0$. We now find that the time delocalization between the two clocks plays a crucial role in the evolution of everything but the clock systems, namely $A,B, A'$ and $B'$. More concretely, the solutions to the constraint equations can again be obtained as follows
\begin{align}
\label{eq:2clock_intervention_state}
    \kket{\Psi}= \frac{1}{2\pi}\int_{\mathds{R}} ds \,dt \,  \varphi(t_2) e^{-is\hat{C}}\ket{0}_{C_1}\ket{t_2}_{C_2}\ket{\phi_0}_S =\frac{1}{2\pi}\int_{\mathds{R}} ds \,dt \, \varphi(t_2)\ket{s}_{C_1}\ket{s+t_2}_{C_2} \mathcal{U}(s,t_2)\ket{\phi_0}_S\, ,
\end{align}
where the unitary evolution of $S$ now explicitly depends on $t_2$ and hence the time delocalization $\varphi(t_2)$, as 
\begin{align}
\label{eq:U_switch}
&\mathcal{U}(s, t_2)\coloneqq
    \begin{cases} 
      e^{-i(s+\frac{t_2}{2})\hat{H}_{S}} & s < \tau_a \text{ and } s < \tau_b-t_2 \\
      e^{-i(s+\frac{1}{2}(t_2-\tau_b))\hat{H}_{S}}U_{BB'}e^{-\frac{i}{2}\tau_b\hat{H}_{S}} & s < \tau_a \text{ and } s > \tau_b-t_2 \\
      e^{-i(s-\frac{1}{2}\tau_a)\hat{H}_{S}}U_{AA'} e^{-\frac{i}{2}(\tau_a+t_2-\tau_b)H_{S}}
      U_{BB'}e^{-\frac{i}{2}\tau_b\hat{H}_{S}} & \tau_b-t_2 <\tau_a < s \\
      e^{-i(s-\frac{1}{2}\tau_a)\hat{H}_{S}} U_{AA'}e^{-\frac{i}{2}(\tau_a+t_2)\hat{H}_{S}} & s > \tau_a \text{ and } s < \tau_b-t_2 \\
      e^{-i(s+\frac{1}{2}(t_2-\tau_b))\hat{H}_{S}}U_{BB'} e^{-\frac{i}{2}(\tau_b- t_2-\tau_a)H_{S}}
      U_{AA'}e^{-\frac{i}{2}(\tau_a+t_2)\hat{H}_{S}} & \tau_a <\tau_b-t_2 < s \, . \\
   \end{cases}
\end{align}
Note that this a priori contains both orders of the operations $\mathcal{M}_A$ and $\mathcal{M}_B$, which will be relevant in our consideration of indefinite order in Sec.~\ref{sec:RelIndefOrder}. The conditional states are given by
\begin{align}
    \ket{\psi_{|C_1}(\tau_1 )} &= \int dt_2 \,\varphi(t_2)\ket{\tau_1+t_2}_{C_2} \mathcal{U}_{|C_1}(\tau_1, t_2)\ket{\phi_0}_S \, , \label{eq:conditional_states_2I2CC1} \\
    \ket{\psi_{|C_2}(\tau_2 )} &= \int dt_2 \,\varphi(t_2)\ket{\tau_2}_{C_2} \mathcal{U}_{|C_2}(\tau_2, t_2)\ket{\phi_0}_S \, ,
    \label{eq:conditional_states_2I2C}
\end{align}
where $\mathcal{U}_{|C_1}(\tau_1, t_2)=\mathcal{U}(\tau_1, t_2)$ and $\mathcal{U}_{|C_2}(\tau_2, t_2)=\mathcal{U}(\tau_2-t_2, t_2)$ with $\mathcal{U}(\tau, t_2)$ given by Eq.~\eqref{eq:U_switch}. The explicit $t_2$-dependence of the conditional evolutions means that now there is some time delocalization of events in both reference frames. To understand this let us first consider the case where $\varphi(t_2)$ is sharply peaked around 0, $\tau_b > \tau_a$ and $\tau_b -\tau_a$ is large enough that the second and third condition in Eq.~\eqref{eq:U_switch} are never fulfilled. In that case, the evolutions with respect to the two clocks are determined by
\begin{align}
\label{eq:condi_evo1}
    &\mathcal{U}_{|C_1}(\tau_1, t_2)=
    \begin{cases} 
      e^{-i(\tau_1+\frac{t_2}{2})\hat{H}_{S}} & \tau_1 < \tau_a \\
      e^{-i(\tau_1-\frac{1}{2}(\tau_a-t_2))\hat{H}_{S}}U_{AA'B'}e^{-\frac{i}{2}\tau_a\hat{H}_{S}} &  \tau_a < \tau_1 < \tau_b-t_2  \\
      e^{-i(\tau_1-\frac{1}{2}\tau_b)\hat{H}_{S}}U_{BB'} e^{-\frac{i}{2}(\tau_b-\tau_a+t_2)H_{S}}
      U_{AA'}e^{-\frac{i}{2}\tau_a\hat{H}_{S}} & \tau_b-t_2 < \tau_1 \, ,
   \end{cases}\\
\label{eq:condi_evo1}
      &\mathcal{U}_{|C_2}(\tau_2, t_2)=
    \begin{cases} 
      e^{-i(\tau_2-\frac{t_2}{2})\hat{H}_{S}} & \tau_2 < \tau_a+t_2 \\
      e^{-i(\tau_1-\frac{1}{2}(\tau_a+t_2))\hat{H}_{S}}U_{AA'}e^{-\frac{i}{2}\tau_a\hat{H}_{S}} &  \tau_a+t_2 < \tau_2 < \tau_b  \\
      e^{-i(\tau_2-t_2-\frac{1}{2}\tau_b)\hat{H}_{S}}U_{BB'} e^{-\frac{i}{2}(\tau_b-\tau_a+t_2)H_{S}}
      U_{AA'}e^{-\frac{i}{2}\tau_a\hat{H}_{S}} &\tau_b< \tau_2 \, ,
   \end{cases}
\end{align}
where according to $C_1$ the application of $U_{BB'}$ is subject to time delocalization, while in the reference frame of $C_2$ the operation given by $U_{AA'}$ appears time-delocalized, cf. Fig.~\ref{fig:causlaity_2clocks}\hyperref[fig:causlaity_2clocks]{(b)}. In each temporal reference frame the intervention timed by the clock of that frame has a sharp time of application, while the operation timed by the other clock appears delocalized in time. Nevertheless, the order of the two operations, as well as the evolution between them, is the same according to both clocks. If this evolution enables signaling between the two subsystem, then $A$ can causally influence $B$. Similarly we can obtain a scenario where $B$ can causally influence $A$, for a different class of constraint equations where $\tau_a>\tau_b$ but $\varphi(t_2)$ is still peaked around 0 sharply enough.

\subsection{Indefinite ordering of quantum operations}\label{sec:RelIndefOrder}

If, on the other hand, $\tau_a$ and $\tau_b$ are closer together, specifically if the non-negligible range of $\varphi(t_2)$ is larger than $|\tau_a-\tau_b|$, then both orders of operations $\mathcal{M}_A$ and $\mathcal{M}_B$ contribute to the evolution in Eq.~\eqref{eq:U_switch} and the causal relations between $A$ and $B$ are no longer well defined. 
Note, that this is categorically different from the delocalization problem encountered in Sec.~\ref{ssec:OperationalCausalityIssues}. There, due to subsystem delocalization operations suitable for investigating causal relations in one temporal frame transformed to operations unsuitable to do so in another frame. Here instead, the time delocalization together with the properties of the constraint operator lead to a scenario where the ordering of the operations is not well defined, in \emph{both temporal frames}. This scenario describes \emph{indefinite causal order} rather than a breakdown of operational causality within relational dynamics. 

\begin{figure}[h]
    \centering
\begin{tikzpicture}[scale=0.55]
\draw[-latex,thick, draw=LimeGreen] (-1.75,-2.5)-- (-1.75,6.75);
\node[] (C)at (-2.5,-2){\color{LimeGreen}$C_1$};

\draw[thick] (0,-1.25) -- (0,6.5);
\draw[thick] (2,-1.25) -- (2,6.5);

\node[] (A)at (0,-2){$ A$};
\node[] (B)at (2,-2){$ B$};
\node[] (C2)at (4,-2){\color{cyan}$ C_2$};

\draw[draw=black,thick,fill=white] (-1,-0.5) rectangle (5.5,5.5);
\node[] (E)at (2.5,2.5){$ {\mathcal{U}_{|C_1}}$};

\draw[thick, color=cyan] (4,-1.25) -- (4,6.5);

\draw[Circle-,LimeGreen]  (-1.85,2.5) -- (0,2.5);
\node[draw=violet, fill=violet!20] (intervention)at (0,2.5){\color{violet}{$\mathcal{M}_A$}};

\node[draw=BlueViolet, fill=BlueViolet!20,semitransparent] (interventionB1)at (2,4.25){\color{BlueViolet}{$\mathcal{M}_B$}};
\draw[Circle-, semitransparent, color=cyan]  (4.15,4.25) -- (interventionB1);

\node[draw=BlueViolet, fill=BlueViolet!20,semitransparent] (interventionB2)at (2,0.5){\color{BlueViolet}{$\mathcal{M}_B$}};
\draw[Circle-, semitransparent, color=cyan]  (4.15,0.5) -- (interventionB2);

\draw[-latex,thick, draw=cyan] (-1.75+10,-2.5)-- (-1.75+10,6.75);
\node[] (C)at (-2.5+10,-2){\color{cyan}$C_2$};

\draw[thick] (0+10,-1.25) -- (0+10,6.5);
\draw[thick] (2+10,-1.25) -- (2+10,6.5);

\node[] (A)at (0+10,-2){$ A$};
\node[] (B)at (2+10,-2){$ B$};
\node[] (C2)at (4+10,-2){\color{LimeGreen}$ C_1$};

\draw[draw=black,thick,fill=white] (-1+10,-0.5) rectangle (5.5+10,5.5);
\node[] (E)at (2.5+10,2.5){$ {\mathcal{U}_{|C_2}}$};

\draw[thick, color=LimeGreen] (4+10,-1.25) -- (4+10,6.5);

\draw[Circle-,cyan]  (-1.85+10,2.5) -- (0+10,2.5);
\node[draw=BlueViolet, fill=BlueViolet!20] (intervention)at (0+10,2.5){\color{BlueViolet}{$\mathcal{M}_B$}};

\node[draw=violet, fill=violet!20,semitransparent] (interventionB1)at (2+10,4.25){\color{violet}{$\mathcal{M}_A$}};
\draw[Circle-, semitransparent, color=LimeGreen]  (4.15+10,4.25) -- (interventionB1);

\node[draw=violet, fill=violet!20,semitransparent] (interventionB2)at (2+10,0.5){\color{violet}{$\mathcal{M}_A$}};
\draw[Circle-, semitransparent, color=LimeGreen]  (4.15+10,0.5) -- (interventionB2);
\end{tikzpicture}

    \caption{The quantum switch within the Page-Wootters formalism: If operations are timed by different clocks, with respect to each clock the operation timed by the other clock appears delocalized in time (see Fig.~\ref{fig:causlaity_2clocks}). We can tailor solutions to the constraint equations such that for $C_1$ the operation timed by $C_2$ appears in a temporal superpositions of being applied before and after the operation timed by $C_1$, and vice versa for the roles of $C_2$ and $C_1$ reversed. The respective conditional states and dynamics correspond to the ``causal reference frame decomposition'' of the quantum switch.}
    \label{fig:causlaity_2clocks_switch}
\end{figure}
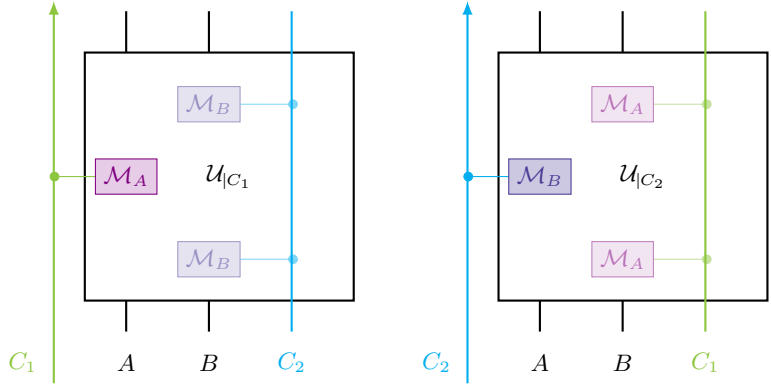

In fact, the constraint operator in Eq.~\eqref{eq:2clock_intervention} allows for describing the evolution of the \emph{quantum switch}~\cite{baumann2022noncausal,castro2020quantum} for a certain choice of kinematical state $\ket{\phi}$, $\tau_a$, $\tau_b$ and $\varphi(t_2)$. Let $\tau_a=\tau_b= \overline{\tau}$, meaning that each clock associates the same time with the application of its respective operation (which does not imply synchronization between them). Moreover, consider the case where $\varphi(t_2)$ is bimodal, with the two peaks symmetric around 0. The conditional evolution according to $C_1$ is then given by Eq.~\eqref{eq:conditional_states_2I2CC1}, with
\begin{align}
\label{eq:U_switch_C1}
&\mathcal{U}_{|C_1}(\tau_1, t_2)=
    \begin{cases} 
      e^{-i(\tau_i+\frac{t_2}{2})\hat{H}_{S}} & \tau_1< \overline{\tau} \text{ and } \tau_1 < \overline{\tau}-t_2 \\
      e^{-i(\tau_i+\frac{1}{2}(t_2-\overline{\tau}))\hat{H}_{S}}U_{BB'}e^{-\frac{i}{2}\overline{\tau}\hat{H}_{S}} & \overline{\tau}-t_2<\tau_1<\overline{\tau} \\
      e^{-i(\tau_i-\frac{1}{2}\overline{\tau})\hat{H}_{S}}U_{AA'} e^{-\frac{i}{2}t_2 H_{S}}
      U_{BB'}e^{-\frac{i}{2}\overline{\tau}\hat{H}_{S}} & \overline{\tau}-t_2 <\overline{\tau} < \tau_1 \\
      e^{-i(\tau_i-\frac{\overline{\tau}}{2})\hat{H}_{S}} U_{AA'}e^{-\frac{i}{2}(\overline{\tau}+t_2)\hat{H}_{S}} & \overline{\tau}<\tau_1<\overline{\tau}-t_2 \\
      e^{-i(\tau_i+\frac{1}{2}(t_2-\overline{\tau}))\hat{H}_{S}}U_{BB'} e^{\frac{i}{2}t_2 \hat{H}_{S}}
      U_{AA'}e^{-\frac{i}{2}(\overline{\tau}+t_2)\hat{H}_{S}} & \overline{\tau}<\overline{\tau}-t_2<\tau_1 \, , \\
   \end{cases}
\end{align}
where the conditions for the appearance of both orderings of $U_{AA'}$ and $U_{BB'}$ are now met. In particular, for the peak of $\varphi(t_2)$ at $t_2>0$, only the first three conditions can be satisfied, and $\mathcal{M}_B$ is applied before $\mathcal{M}_A$. For the peak at $t_2<0$, on the other hand, only the first one and last two conditions for the evolution can be fulfilled, and $\mathcal{M}_A$ is therefore applied before $\mathcal{M}_B$. As in Sec.~\ref{ssec:RelationalCausality &multiple Clocks}, the application of $U_{BB'}$ appears time-delocalized in the frame of clock $C_1$, with the extent of this delocalization determined by the finite extent of the two peaks of $\varphi(t_2)$. One obtains an analogous expression for $\mathcal{U}_{|C_2}(\tau_2, t_2)$ containing both orders of operations but where the application of $U_{AA'}$ is delocalized in time. Given the assumed symmetry of $\varphi(t_2)$, these conditional evolutions describe the ``causal reference frame decomposition'' of the quantum switch (see~\cite{guerin2018observer,baumann2022noncausal}), as illustrated in Fig.~\ref{fig:causlaity_2clocks_switch}.

\section{Discussion and conclusion}
\label{sec:Discussion}

In this work we have investigated the modeling of interventions and measurements in the Page-Wootters formalism, and how the resulting understanding of causal relations depends upon the choice of clock, in particular the notions of time-locality and subsystem-locality according to those clocks. First we have elucidated a necessary temporal delocalization when comparing different frames, though in the absence of other factors (for example certain clock-system interactions~\cite{castro2020quantum}) one may consider this delocalization to be arbitrarily small for ideal clocks. This exemplifies the necessity of explicitly considering the form of correlations between clocks when comparing them as choices of reference frame, a feature which is absent in the abstract changes of ``external'' frame usually encountered in physics. To avoid a potential source of confusion, we stress that the delocalization discussed here is different in nature from the time-nonlocal evolution that can result from clock-system interactions~\cite{Smith2019}.

In light of the fact that two clocks may disagree not only on the specific time of an event, but indeed whether it occurs at a well-defined time at all, we examined the possibility of two clocks disagreeing on the direction of time, showing that this may only occur in an approximate sense. Under this approximation, we found that if those clocks are to disagree about the order of two spontaneous processes (i.e. not triggered by interactions with the clocks), then those processes must be time-symmetric.

Moving on to the relational modeling of interventions, we began with an approach whereby, for example, the difference between the occurrence or not of some intervention was understood as a comparison between two different solutions to the same constraint equation. In this manner, the counterfactual reasoning used to determine causal relations is incorporated as a comparison of different ``alternate histories'': one in which the intervention occurred, and one in which it did not. We found that, while this is workable in a particular reference frame, the relativity of subsystems~\cite{ali2022quantum} means that the corresponding state-comparison in a different frame does not allow for an interpretation in terms of causal influences from one party to another.

In order to obtain a picture where causal relations are independent of the chosen clock, one must therefore incorporate the interventions at the level of the constraint operator. Adapting methods from~\cite{Smith2019,castro2020quantum}, we have shown how this results in the preservation of the subsystem locality of an intervention or measurement across perspectives. However the extent of its joint localizablity in time is determined by the specific form of correlations between the clocks considered. For states where the necessary temporal delocalization between clocks is negligible, interventions are approximately time-local in both frames, in addition to being subsystem-local. If the temporal delocalization is small (compared to the timescales of the other processes) but non-negligible, however, one finds some temporal smearing of the relevant processes in one or another frame. 

Considering then the more extreme case of delocalization between clock states, one can obtain a situation where the ordering of some operations is indefinite, and describe a relational version of the quantum switch in this regime. Note that, while each clock may disagree about the particulars, they do not disagree that the order of operations is indefinite, in keeping with~\cite{de2025indefinite}. Further note that, in the relational picture here, the quantum switch arose in part due to a tailored choice of physical state (via a corresponding kinematical state), where the time delocalization $\varphi(t_2)$ plays the role of the control. In~\cite{castro2020quantum}, on the other hand, time delocalization between the two clocks is brought about by gravitational interaction with a mass in spatial superposition, which constitutes an actual control system. Hence, the scenario in~\cite{castro2020quantum} describes a ``physical'' quantum switch, while the relational evolution discussed in Sec.~\ref{sec:RelIndefOrder} describes how a certain choice of ``history'' (i.e. physical state) allows for this type of indefinite causal order.

The common assumption of ideal clocks, which have the unphysical property that their energy spectrum is unbounded below~\cite{pauli1958allgemeinen,garrisonCanonicallyConjugatePairs1970}, was made throughout this paper except where otherwise stated. If this assumption were relaxed, the ``interaction-picture'' method described in Appendix~\ref{app:Interaction terms for timed interventions} would no longer be applicable, at least in its current form, and it is unclear exactly how the conclusions above would change.  Indeed in~\cite{hausmann2025measurement}, the authors showed how a measurement described with respect to a non-ideal clock leads to another kind of temporal indefiniteness, an effect which we have not considered here. Moreover, outside of the ideal limit, tensor factorizability itself can become frame-dependent~\cite{ali2022quantum}, and thus care must be taken in the relationship between the chosen clock and subsystems in question in order to avoid this. 

In addition to clock ideality, Appendix~\ref{app:Interaction terms for timed interventions} assumed the clock-system couplings to be effectively instantaneous, and relaxing this assumption to account for the necessary duration of the interventions would likewise affect the results above. While it seems that the retention of the subsystem-local nature of interventions between frames was not contingent upon this assumption, the question of time-locality of interventions in a given frame must be reexamined. Furthermore, in the consideration of time-delocalization in Secs.~\ref{sec:interventions&PW}, we have ignored the potential for overlap between the delocalized interventions and measurements and the dynamics represented by maps of the form $\mathcal{E}_{\vert C}$, and more refined consideration of this aspect could yield further complications to the modeling of operational causality in a relational setting.

While, in the absence of interactions, the Page-Wootters formalism is one of an equivalent set of approaches to relational dynamics~\cite{Trinity2021,hohn2021equivalence,chataignier2026relational}, it is unclear to what extent this equivalence is affected by non-zero interactions. If the equilvalence turns out to be broken in this case, then the conclusions obtained here may differ from what would be obtained in the picture of relational Dirac observables\cite{rovelli1990quantum,rovelli1991quantum,rovelli1991time,rovelliQuantumGravity2004,thiemannModernCanonicalQuantum2008}, or in the the relational Heisenberg picture obtained via quantum deparameterization~\cite{Hoehn:2018whn,hohn2020switch}.

Our results exemplify the necessity of a careful consideration of the joint localizability of clocks, and in particular its limitations. They moreover suggest that operational causality in relational quantum dynamics can be formulated consistently only if interventions and measurements are treated as part of the dynamics rather than via ``initial conditions'' and abstract operations respectively. The role of joint temporal localizability in such a formulation, together with the results of~\cite{castro2020quantum,hausmann2025measurement}, points toward a broader picture in which temporal delocalization is a generic feature in relational quantum settings, and we have seen here how this permits relational realizations of indefinite causal order. At a more fundamental level, there is the question of how synchronization among many distributed clocks can arise under the limited joint temporal localizability discussed here, and with it how a more familiar notion of reference frame, closer to that envisaged in special relativity~\cite{einstein1905elektrodynamik}, and a corresponding notion of causality may emerge.

\section*{Acknowledgments}

The authors thank V. Vilasini and Carlo Cepollaro for helpful discussions. V.B. is funded by the Austrian Science Fund (FWF) grant No.\ ESP 520-N. M.~P.~E.~L This research was funded in whole or in part by the Austrian Science Fund (FWF) 10.55776/I6949, 10.55776/COE1 and the European Union – NextGenerationEU

\bibliography{refs.bib}

\clearpage

\begin{appendix}

\section{Frame-dependent ordering of operations}
\label{app:frame dependent ordering}

Here, we discuss the very limited conditions under which the clock-dependent time direction in Sec.~\ref{sec:time-reversal} can be understood as a temporal-reference-frame dependent ordering of events. In order to be able to consider general CP maps, we must introduce ancilla systems. We thus consider a system $S$ consisting of three subsystems, $\mathcal{H}_{S}=\mathcal{H}_{S^{\prime}} \otimes \mathcal{H}_{A^{\prime}}\otimes \mathcal{H}_{B^{\prime}}$, i.e. a target system $S^{\prime}$ and an ancilla system $A^{\prime}$ for the Stinespring dilation of a CP map $\mathcal{E}_A$ on $S^{\prime}$, and an ancilla system $B^{\prime}$ for the Stinespring dilation of a CP map $\mathcal{E}_B$ on $S^{\prime}$. The state $\ket{\phi_0}$ in Eqs.~\eqref{example_state}-\eqref{eq:initial state example} now includes the ancilla systems, and we choose $\ket{\phi_0}_S=\ket{\theta_0}_{S^{\prime}}\ket{a_0}_{A^{\prime}}\ket{b_0}_{B^{\prime}}$. With this setup, we can then phrase our question as follows: since the clocks run in different directions, are there circumstances in which one sees the system $S^{\prime}$ undergoing $\mathcal{E}_A$ and then $\mathcal{E}_B$, while the other sees it undergoing to $\mathcal{E}_B$ and then $\mathcal{E}_A$? 

Analogous to Eqs.~\eqref{Perspectival_states1} and~\eqref{Perspectival_states2}, we thus obtain the states of $S^{\prime}$ relative to each clock:
\begin{align}
\rho_{S^{\prime}|C_1} (\tau_1) &:=  \text{Tr}_{C_2,A^{\prime},B^{\prime}} \left( \ket{\psi_{| C_1}(\tau_1)}\bra{\psi_{| C_1}(\tau_1)} \right) \nonumber \\
&= \sumint_{\,a,b} \bra{a,b} e^{-i\tau_1 \hat{H}_S} \left(\proj{\theta_0}\otimes \proj{a_0}\otimes\proj{b_0}\right) e^{i\tau_1 \hat{H}_S} \ket{a,b}\, ,
\label{Perspectival_states1ex}
 \\
\rho_{S^{\prime}|C_2} (\tau_2) &:= \text{Tr}_{C_1,A^{\prime},B^{\prime}} \left( \ket{\psi_{| C_2}(\tau_2)}\bra{\psi_{| C_2}(\tau_2)} \right) \nonumber \\
&= \sumint_{\,a,b} \bra{a,b} \int dt |\varphi(t)|^2 e^{i(\tau_2-t) \hat{H}_S} \left(\proj{\theta_0}\otimes \proj{a_0}\otimes\proj{b_0}\right) e^{-i(\tau_2-t) \hat{H}_S} \ket{a,b}\, ,
\label{Perspectival_states2ex}
\end{align}
where $a$ and $b$ label arbitrary bases of the Hilbert space associated with the ancillary systems $A'$ and $B'$ respectively. Now, given the necessarily continuous action of the Hamiltonian $\hat{H}_S$ in both cases, we see that an issue immediately arises. Let us focus for now on $C_1$'s perspective. If Eq.~\eqref{Perspectival_states1ex} should represent the composed Stinespring dilation of the maps $\mathcal{E}_A$ and $\mathcal{E}_B$, then there must exist some $\Delta \tau$ such that $e^{-i\Delta \tau\hat{H}_S}=U_B U_A$, where $U_A$ ($U_B$) is the unitary dilating $\mathcal{E}_A$ ($\mathcal{E}_B$), and thus coupling $S'$ with $A'$ ($B'$). However, the necessary coupling terms in $\hat{H}_S$ act at all times, and cannot a priori be neatly separated into two ordered times.

In the case where $\mathcal{E}_A$ and $\mathcal{E}_B$ are themselves unitary, this is not problematic. The ancillas $A'$ and $B'$ are unnecessary and the required Hamiltonian $\hat{H}_S$ is simply given (up to an arbitrary constant) by the operator whose exponential is equal to the product of the unitaries in question. If we wish to model more general CP maps, on the other hand, we must either explicitly include a mechanism for timing one map and then the other (see Sec.~\ref{sec:interventions&PW}), or ``fine-tune'' the input states $\ket{a_0}_{A^{\prime}}$ and $\ket{b_0}_{B^{\prime}}$ of the ancillas, and their coupling to $S'$, such that we arrive at a situation comparable to sequential scattering events. In other words, they are chosen such that for some part of the evolution in Eq.~\eqref{Perspectival_states1ex}, there is effectively no interaction between $B'$ and the other systems, and then for some later part of the evolution, this is no longer the case, and the interaction between $A'$ and the other systems is negligible. Assuming that such a choice is possible (and leaving aside the question of when it is), then there exists $\Delta \tau$ such that $e^{-i\Delta \tau\hat{H}_S}\approx U_B U_A$ and furthermore
\begin{align}
\rho_{S^{\prime}|C_1}(\Delta \tau)\approx \tr_{A' B'C_2}( U_B U_A\proj{\theta_0}_{S^{\prime}} U_A^{\dagger} U_B^{\dagger} )  \approx \mathcal{E}_B\big[\mathcal{E}_A [\proj{\theta_0}_{S^{\prime}}]\big].
\label{eq:perspective_dep_ordering1}
\end{align}
Considering now the perspective of the second clock, $C_2$, and assuming that the two clocks are approximately synchronized, i.e. that $\varphi(t)$ is tightly-peaked around $t=0$, such that $|\varphi(t)|^2\approx\delta(t)$, we then have
\begin{align}
\rho_{S^{\prime}|C_2}(\Delta \tau)\approx  \tr_{A' B'C_2}(U_A^{\dagger} U_B^{\dagger} \proj{\theta_0}_{S^{\prime}} U_B U_A) \neq \mathcal{E}_A[\mathcal{E}_B [\proj{\theta_0}]]\, .
\label{eq:perspective_dep_ordering2}
\end{align}
We thus see that, even given the approximations and assumption above, we still don't arrive at an example of perspective-dependent ordering of CP maps, except in the case where their Steinspring dilations are time-reversal symmetric, i.e.~where $U_A$ and $U_B$ are Hermitian. Examples of operations which are both unitary and Hermitian include the Pauli operators on a qubit.

\section{Example of operational causality}
\label{app:example}

Here, we present a simple example of two qubits and discuss how to investigate operational causality as depicted in Fig.~\ref{fig: signalling} in the main text. We then apply our approaches for incorporating operational causality into the Page-Wootters formalism from Secs.~\ref{sec:OperationalCausality&PW} and~\ref{sec:interventions&PW} to this example. Let us consider the following two Hamiltonians
\begin{align}
\label{eq:H_AB-noint}
    &\hat{H}_{S}= X_A\otimes\mathds{1}+\mathds{1}\otimes X_B, \\
\label{eq:H_AB-int}
    &\hat{H}'_{S}= X_A\otimes X_B, 
\end{align}
where $X$ is the Pauli-x-matrix, together with the initial two-qubit state
\begin{align}
    \rho(t_i)= \proj{0}_A\otimes\proj{0}_B.
    \label{eq:example-intial}
\end{align}
While $\hat{H}_{S}$ describes the two subsystems $A$ and $B$ evolving independently, $\hat{H}'_{S}$ corresponds to the two qubits interacting. The respective unitary evolutions are
\begin{align}
\label{eq:U_AB-noint}
    & U_{S}= \cos^2(t) \mathds{1}-i\cos(t)\sin(t)(X_A\otimes\mathds{1}+\mathds{1}\otimes X_B)-\sin^2(t) X_A\otimes X_B \,, \\
\label{eq:U_AB-int}
    & U'_{S}= \cos(t) \mathds{1}-i\sin(t) X_A\otimes X_B\, .
\end{align}
In order to determine the causal relations between $A$ and $B$, we have to apply a quantum operation to one of the subsystems, say A, at time $t_i$ and determine whether the choice of operation has a detectable influence on the other subsystem, $B$, at a later time $t_f>t_i$. Such an intervention on subsystem $A$ corresponds to replacing the initial state in Eq.~\eqref{eq:example-intial} by
\begin{align}
    \overline{\rho}(t_i)= \rho_A\otimes\proj{0}_S,
    \label{eq:example-intervention}
\end{align}
where $\rho_A= \mathcal{M}_A (\proj{0})$ depends on which operation is performed. This intervention on $A$ can be detected on subsystem $B$ at $t_f$ if 
\begin{align}
    \label{eq:check_signaling}
    \overline{\rho}^{(\prime)}_B(t_f) = \tr_A\left(U^{(\prime)}_{S}(t_f)\overline{\rho}(t_i)U^{(\prime)\dagger}_{S}(t_f)\right) \neq 
    \tr_A \left( 
    U^{(\prime)}_{S}(t_f)\rho(t_i)U^{(\prime)\dagger}_{S}(t_f)
    \right)=\rho^{(\prime)}_B(t_f),
\end{align}
where $\overline{\rho}(t_i)$ and $\rho(t_i)$ are given by Eqs.~\eqref{eq:example-intervention} and~\eqref{eq:example-intial} respectively. The inequality in Eq.~\eqref{eq:check_signaling} means that there exists a quantum operation that can be performed on $B$ which is affected by $A$'s operation, e.g. a measurement at time $t_f$ the outcomes of which are different depending on whether $\mathcal{M}_A$ is applied or not. The evolution $U_{S}$ in Eq.~\eqref{eq:U_AB-noint} leads to  
\begin{align}
\label{eq:rho_bar}
\rho_B(t_f)=\cos^2(t_f)\proj{0}+\sin^2(t_f)\proj{1}+i\cos(t_f)\sin(t_f)(\ketbra{0}{1}-\ketbra{1}{0})=\overline{\rho}_B(t_f),
\end{align}
which in independent of $\rho_A$ and hence any operation performed on $A$. This means that there is no causal influence from $A$ to $B$ as expected since in this case the two qubits evolve independently.
Conversely, for $U'_{S}$ in Eq.~\eqref{eq:U_AB-int} we obtain
\begin{align}
    \rho'_B(t_f)&=\cos^2(t_f)\proj{0}+\sin^2(t_f)\proj{1} \\
    \label{eq:rho'_bar}
    &\neq \cos^2(t_f)\proj{0}+\sin^2(t_f)\proj{1}+i\cos(t_f)\sin(t_f) \tr(X\rho_A)(\ketbra{0}{1}-\ketbra{1}{0})=\overline{\rho}'_B(t_f)
\end{align}
which means that $A$ can signal to $B$ if their operations are chosen accordingly. More explicitly consider, for example, the intervention $\mathcal{M}_A$ at $t_i$ being a Hadamard-gate, i.e. $\overline{\rho}'(t_i)=M_A  \proj{0}M_A^{\dagger}\otimes \proj{0}_B$ with $M_A=\frac{1}{\sqrt{2}}\begin{pmatrix}
1 & 1 \\
1 & -1 
\end{pmatrix}$, and the operation $\mathcal{O}_B$ at $t_f$ is a measurement in the Pauli-y basis, i.e. $O_B=\proj{+y}-\proj{-y}$. Then, we find that
\begin{align}
\label{eq:p-example1}
    p(+y|\text{no intervention})&=\bra{+y}\rho'_B(t_f)\ket{+y}=\frac{1}{2}=p(-y|\text{no intervention})\\
\label{eq:p-example2}
    p(+y|\text{intervention})&= \bra{+y}\overline{\rho}'_B(t_f)\ket{+y}=\frac{1}{2}\big( 1-\sin(2t_f)\big)\\
\label{eq:p-example3}
    p(-y|\text{intervention})&= \bra{-y}\overline{\rho}'_B(t_f)\ket{-y}=\frac{1}{2}\big( 1+\sin(2t_f)\big),
\end{align}
showing that the application of the Hadamard-gate on $A$ at time $t_i$ can be detected by this measurement on $B$ at an appropriately chosen time $t_f$.\\

Now we consider this example within the Page-Wootters formalism as described in Sec.~\ref{ssec:OperationalCausalityIssues} starting with constraint operator(s) 
\begin{align}
    \hat{C}=\hat{H}_C+\hat{H^{(\prime)}}_{S},
\end{align}
where $\hat{H^{(\prime)}}_{S}$ are those in Eqs.~\eqref{eq:H_AB-noint} and~\eqref{eq:H_AB-int} respectively. Solutions to the constraint equation can be obtained from the kinematical state $\ket{\phi}=\ket{t_i}_C\ket{0}_{A}\ket{0}_{B}$ via
\begin{align}
\label{eq:state_1C}
    \kket{\Psi}=\frac{1}{2\pi}\int ds \, e^{-is\hat{C}} \ket{t_i}_C\ket{0}_{A}\ket{0}_{B}
    =\frac{1}{2\pi}\int ds \, \ket{t_i+s}_C e^{-is\hat{H}_{S}^{(\prime)}}\ket{0}_{A}\ket{0}_{B} \, ,
\end{align}
which allows for defining the following conditional states for different interventions $\mathcal{M}_A$ 
\begin{align}
    \ket{\psi^{A}_{|C}(\tau)}=e^{-i(\tau-t_i)\hat{H_{S}^{(\prime)}}} \mathcal{M}_A\ket{0}_A\ket{0}_B \; ,
\end{align}
from which we can reproduce the example above in a straight forward manner. We have $\rho(t_i)=\proj{\psi_{|C}(t_i)}=\proj{0}_A \otimes \proj{0}_B$ and $\overline{\rho}(t_i)=\proj{\psi^{A}_{|C}(t_i)}=\mathcal{M}_A[\proj{0}]_A \otimes \proj{0}_B$ and need to compare
\begin{align}
    \rho_B(t_f)=&\tr_A\left(\proj{\psi_{|C}(t_f)}\right)\\
    &\qquad  \text{ and } \nonumber \\
     \overline{\rho}_B(t_f)=&\tr_A\left(\proj{\psi^{A}_{|C}(t_f)}\right),
\end{align}
which are the states in Eqs.~\eqref{eq:rho_bar} and~\eqref{eq:rho'_bar} respectively, since $\exp(-i\tau\hat{H_{S}^{(\prime)}})=U^{(\prime)}_{S}$.

Note that the Hamiltonians $\hat{H}^{(\prime)}_{S}$ in this example are symmetric in $A$ and $B$ and  we can equally consider  
\begin{align}
    \ket{\psi^{B}_{|C}(\tau)}=e^{-i(\tau-t_i)\hat{H}_{S}^{(\prime)}} \ket{0}_A \mathcal{M}_B\ket{0}_B,
\end{align}
and compare $\tr_B \left( \proj{\psi^{B}_{|C}(t_f)}\right)$ to find signaling from $B$ to $A$ for the case where the two qubits interact, i.e. $\hat{C}=\hat{H}_C+\hat{H}'_{S}$. In fact, since signaling from $A$ to $B$ implies signaling from $B$ to $A$ for any bipartite unitary~\cite{beckman2001causal} we would have found the same for non-symmetric Hamiltonians. In the absence of further subsystems, unitary evolution allowing for a causal influence from $A$ on $B$ equaly allows for $B$ to causally influence $A$.\\ 

Let us now extend this example to a relational scenario with two clock systems corresponding to constraint operator
\begin{align}
    \label{eq:constraint2clock_ex}
    \hat{C}=\hat{H}_{C_1}+\hat{H}_{C_2}+\hat{H}^{(\prime)}_{S} ,
\end{align}
where $\hat{H}^{(\prime)}_{S}$ are again given by Eqs.~\eqref{eq:H_AB-noint} and~\eqref{eq:H_AB-int}. We can construct a physical state as follows
\begin{align}
    \label{eq:2clock_psi phys}
    \kket{\Psi}&= \frac{1}{2\pi}\int ds \, e^{-is\hat{C}} \int dt\, \ket{t_i}_{C_1}\varphi(t) \ket{t}_{C_2}\ket{0}_{A}\ket{0}_B ,\\
    &=\int ds\, dt \, \varphi(t)\ket{t_i+s}_{C_1}\ket{t+s}_{C_2}e^{-is\hat{H}^{(\prime)}_{S}}\ket{0}_A\ket{0}_B ,\nonumber
\end{align}
where we introduced the necessary time delocalization on clock $C_2$ and used the same notation as before. From the perspective of clock $C_1$, we obtain 
\begin{align}
    \ket{\psi^A_{|C_1}(\tau_1)}&=\int dt \, \varphi(t) \ket{t+\tau_1-t_i}_{C_2} U^{(\prime)}_{S}(\tau_1-t_i)\mathcal{M}_A\ket{0}_A\ket{0}_B, \\
    \mathcal{M}_A&= M_A \otimes \mathds{1}_B,\\
    \mathcal{O}_B&=\mathds{1}_A \otimes O_B,
\end{align}
which gives the following probabilities (and signaling behaviors) for measuring $\mathcal{O}_B$, i.e. $O_B=\sum_b b \,\Pi^b$ at time $t_f$.
\begin{align}
    p_{|C_1}(b|\text{intervention})&=\tr\left(\mathds{1}_A\otimes \Pi^b_B \proj{\psi^{A}_{|C_1}(t_f)}\right)\\
    &=\int dt\, |\varphi(t)|^2 \bra{0,0}M^{\dagger}_A  U^{(\prime)}_{S}(t_i-t_f) \Pi^b_B U^{(\prime)}_{S}(t_f-t_i) M_A \ket{0,0},
\end{align}
where $b$ is the result obtained. If we again consider the Pauli-y measurement, i.e. $O_B=\proj{+y}-\proj{-y}$, and compare $M_0=\mathds{1}$ and $M_A=\frac{1}{\sqrt{2}}\begin{pmatrix}
1 & 1 \\
1 & -1 
\end{pmatrix}$ we recover $p(b|\text{no intervention})$ and $p(b|\text{intervention})$ from Eqs.~\eqref{eq:p-example1}-\eqref{eq:p-example3}.
In the reference frame of clock $C_2$, however, we have
\begin{align}
    \ket{\psi^A_{|C_2}(\tau_2)}&=\int dt \, \varphi(t) \ket{\tau_2-t+t_i}_{C_2} U^{(\prime)}_{S}(\tau_2-t+t_i)\mathcal{M}_A\ket{0}_A\ket{0}_B ,\\
    \mathcal{M'}_A&=\int d\alpha \, \proj{\alpha}_{C_1}\otimes e^{-i\alpha\hat{H}^{(\prime)}_{S}} (M_A \otimes \mathds{1}_B) e^{i\alpha \hat{H}^{(\prime)}_{S}} ,\\
    \mathcal{O'}_B&=\int d\alpha \, \proj{\tau_F + \alpha}_{C_1}\otimes e^{-i\alpha\hat{H}^{(\prime)}_{S}}(\mathds{1}_A \otimes O_B) e^{i\alpha \hat{H}^{(\prime)}_{S}}\, .
\end{align}
While the probabilities remain unchanged under the change of clock perspective
\begin{align}
    p_{|C_1}(b|\text{(no) intervention})&= p_{|C_2}(b|\text{(no) intervention}) =
    p(b|\text{(no) intervention})\, ,\nonumber
\end{align}
the operations $\mathcal{M}'_A$ and $\mathcal{O}'_B$ now no longer act on single subsystems (see Fig.~\ref{fig: PW_intervention} in the main text). More concretely, we find 
\begin{align}
    \mathcal{M'}_A&=\int d\alpha\, \proj{\alpha}_{C_1}\otimes
    \left(\cos^2(\alpha) M_A
    +\sin^2(\alpha) XM_AX
    +i \cos(\alpha)\sin(\alpha) [M_A,X]
    \right)\otimes\mathds{1}_B ,
    \\
    \mathcal{O'}_B&=\int d\alpha\, \sum_b \proj{\alpha}_{C_1}\otimes \mathds{1}_A\otimes
    \left(\cos^2(\alpha) \Pi^b_B
    +\sin^2(\alpha) X\Pi^b_BX
    +i \cos(\alpha)\sin(\alpha) [\Pi^b_B,X]
    \right) ,
\end{align}
for the case where $A$ and $B$ do not interact, i.e. $\hat{H}_{S}$. Since here $\mathcal{M'}_A$ acts trivially on $B$, $\mathcal{O'}_B$ acts trivially on $A$ and the probabilities are non-signaling, we can still conclude that there is no causal connection between $A$ and $B$. However, when the two qubits interact, i.e. for $\hat{H}^{\prime}_{S}$, we have 
\begin{align}
    \mathcal{M'}_A&=\int d\alpha\, \proj{\alpha}_{C_1}\otimes
    \left(\cos^2(\alpha) M_A\otimes\mathds{1}_B
    +\sin^2(\alpha) XM_AX\otimes\mathds{1}_B
    +i \cos(\alpha)\sin(\alpha) [M_A,X]\otimes X_B
    \right) ,\\
    \mathcal{O'}_B&=\int d\alpha\, \sum_b b \proj{\alpha}_{C_1}\otimes
    \left(\cos^2(\alpha) \mathds{1}_A \otimes \Pi^b_B
    +\sin^2(\alpha) \mathds{1}_A \otimes X\Pi^b_BX
    +i \cos(\alpha)\sin(\alpha) X_A \otimes [\Pi^b_B,X]
    \right) ,
\end{align}
meaning that both operations jointly act on all the subsystems, $A$, $B$ and $C_1$. This means that in the reference frame of $C_2$ the signaling probabilities can no longer be attributed to a causal influence form $A$ to $B$. 

Finally, we include the interventions above into the constraint Hamiltonian and show how it gives consistent causal reasoning for all reference frames, as discussed in Sec.~\ref{sec:interventions&PW} of the main text. Let us consider constraint Hamiltonians 
\begin{align}
    \label{eq:constraint2clock_ex_interventions}
    \hat{C}=\hat{H}_{C_1}+\hat{H}_{C_2}+\hat{H}^{(\prime)}_{S}+\delta(\hat{T}_1-t_i)\hat{K}_A + \delta(\hat{T}_1-t_f)\hat{K}_{BB'} \, ,
\end{align}
where both interventions are timed by clock $C_1$. When considering the example above we need to compare $\hat{K}_A=0$ corresponding to no intervention to $\hat{K}_A=\frac{\pi}{2\sqrt{2}}(X+Z)$ describing the application of the Hadamard-gate at time $t_i$. The operation at $t_f$ is the van-Neumann measurement in the Pauli-y basis, i.e. $e^{-iK_B}\ket{\pm y}_B\ket{b_0}_{B'}=\ket{\pm y}_B\ket{\pm y}_{B'}$. Since $t_i<t_f$ we obtain
\begin{align}
\label{eq:Psi_phys_intervention-example}
   \kket{\Psi}&= \frac{1}{2\pi} \int_{-\infty}^{t_i} dt_1 \, \int dt_2 \, \varphi(t_2) \ket{t_1}_{C_1}\ket{t_1+t_2}_{C_2} U^{(\prime)}_{S}(t_1)\ket{0,0}_{AB}\ket{a_0}_{A'
}\ket{b_0}_{B'}\\
   &+\frac{1}{2\pi} \int_{\tau_i}^{\tau_f} dt_1 \,  \int dt_2 \, \varphi(t_2) \ket{t_1}_{C_1}\ket{t_1+t_2}_{C_2} U^{(\prime)}_{S}(t_1-t_i)e^{-i\hat{K}_A}U^{(\prime)}_{S}(t_i)\ket{0,0}_{AB}\ket{a_0}_{A'
}\ket{b_0}_{B'} \nonumber\\
    &+\frac{1}{2\pi} \int_{\tau_f}^{\infty} dt_1 \, \int dt_2 \, \varphi(t_2) \ket{t_1}_{C_1}\ket{t_1+t_2}_{C_2} U^{(\prime)}_{S}(t_1-t_f)e^{-i \hat{K}_B} U^{(\prime)}_{S}(t_f-t_i)e^{-i\hat{K}_A}U^{(\prime)}_{S}(t_i)\ket{0,0}_{AB}\ket{a_0}_{A'
}\ket{b_0}_{B'}
    ,\nonumber
\end{align}
which gives the conditional evolution from Eq.~\eqref{eq:int_1clock_state} in the main text and the probabilities form Eqs.~\eqref{eq:p-example1}-~\eqref{eq:p-example3} provided that $\varphi(t_2)$ is peaked around $0$ sharply enough, compare Sec.~\ref{ssec:RelationalCausality &multiple Clocks}. We have, hence fully incorporated the above example of operational causality into the Page-Wootters formalism via including timed interventions into the constraint operator.

\section{Timed interventions in the Page-Wootters formalism}
\label{app:Interaction terms for timed interventions}

Here, we show how tailored interaction terms in the constraint operator give rise to physical states with timed interventions. We start with a general form and transform it in a way that allows us to evaluate $e^{-i\alpha\hat{C}}$. We then consider the limiting case of instantaneous interventions.

\subsection{An interaction picture for clock-conditioned operations}

Let $\mathcal{H}_\mathrm{kin}=\mathcal{H}_{C}\otimes\mathcal{H}_{S}$, and let $\hat{V}(\hat{T})$ denote a self-adjoint operator on $\mathcal{H}_\mathrm{kin}$ of the form
\begin{equation}
    \hat{V}(\hat{T}) := \int_\mathds{R}dt\,f(t) \ketbra{t}{t}\otimes\hat{K}_{S} .
    \label{eq:H_int}
\end{equation}
Consider the Hamiltonian constraint $\hat{C}\kket{\Psi}=0$ with
\begin{equation} \label{eAppConstr}
    \hat{C}= \hat{H}_C +\hat{H}_S +  \hat{V}(\hat{T}).
\end{equation}
One can then move to an ``interaction picture'' as follows~\cite{castro2025private}. Defining $U_\text{int}:= e^{i\hat{T}\otimes\hat{H}_{S}}$, and $\hat{C}_\text{int}:= U_\text{int}\hat{C}U_\text{int}^\dag$, and then the constraint equation $\hat{C}\kket{\Psi}=0$ can be rewritten as
\begin{equation}
	    \hat{C}\kket{\Psi} = U_\text{int}\hat{C}\kket{\Psi} = U_\text{int}\hat{C}U_\text{int}^\dag U_\text{int}\kket{\Psi} \equiv \hat{C}_\text{int}\kket{\Psi_\text{int}}=0,
\end{equation}
where $\kket{\Psi_\text{int}}:= U_\text{int}\kket{\Psi}$. We thus have a new set of physical states $\lbrace\kket{\Psi_\text{int}}\,:\,\hat{C}_\text{int}\kket{\Psi_\text{int}}=0\rbrace$, which are unitarily equivalent to the original set. On can then construct a projector from any element of the kinematical Hilbert space $\ket{\phi}$ to the corresponding ``interaction picture'' physical state in an analogous manner to Eq.~\eqref{eq:physical_state_from_deltaC}, i.e.
\begin{equation}
	\kket{\Psi_\text{int}}=\delta(\hat{C}_\text{int})\ket{\phi}= \frac{1}{2\pi} \int_\mathds{R} ds\, e^{-is\hat{C}_\text{int}}\ket{\phi}
\end{equation}

The form of $\hat{C}_\text{int}$ can be obtained via the Baker–Campbell–Hausdorff formula:
\begin{equation} \label{eIntConstraint}
    \hat{C}_\text{int} = \hat{H}_{C} + \hat{V}_\mathrm{int}(\hat{T}),
\end{equation}
where
\begin{equation}
    \hat{V}_\mathrm{int}(\hat{T}) := U_\text{int}\hat{V}(\hat{T})U_\text{int}^\dag = \int_\mathds{R}dt\,f(t) \ketbra{t}{t}\otimes e^{it\hat{H}_{S}}\hat{K}_{S} e^{-it\hat{H}_{S}} \equiv \int_\mathds{R}dt\,f(t) \ketbra{t}{t}\otimes \hat{K}_{S}(t),
\end{equation}
with $\hat{K}_{S}(t):= e^{it\hat{H}_{S}}\hat{K}_{S} e^{-it\hat{H}_{S}}$.

\subsection{Exponentiating the interacting constraint Hamiltonian}

Now, note that we can use the ``interaction picture'' to write
\begin{equation}\label{eInteractionDelta}
    e^{-i\alpha \hat{C}} = U_\text{int}^\dag  e^{-i\alpha \hat{C}_\mathrm{int}} U_\text{int} .
\end{equation}
We now follow the Supplemental Material of Ref.~\cite{castro2020quantum} to obtain an expression of the middle factor in terms of an time-ordered exponential. First, using Eq.~\eqref{eIntConstraint} and the Trotter-Suzuki formula, we have
\begin{equation} \label{eTrotter}
    e^{-i\alpha \hat{C}_\mathrm{int}} = \lim_{N\to\infty} \left(e^{-i\frac{\alpha}{N}\hat{H}_{C}} e^{-i\frac{\alpha}{N} \hat{V}_\mathrm{int}(\hat{T})} \right)^{N} .
\end{equation}
Recalling that the eigenstates $\lbrace\ket{t}\rbrace_{t\in\mathds{R}}$ of $\hat{T}$ form a basis of $\mathcal{H}_{C}$, and taking the tensor product with an arbitrary basis $\lbrace\ket{r}\rbrace_{r}$ of $\mathcal{H}_{S}$, one can consider the action of the base on the right-hand side of Eq.~\eqref{eTrotter} on an arbitrary basis element $\ket{t}_{C}\ket{r}_{S}$ of $\mathcal{H}_\mathrm{kin}$:
\begin{equation}
    e^{-i\frac{\alpha}{N}\hat{H}_{C}} e^{-i\frac{\alpha}{N} \hat{V}_\mathrm{int}(\hat{T})} \ket{t}_{C}\ket{r}_{S} = e^{-i\frac{\alpha}{N}  f(t) \hat{K}_{S}(t)} \ket{t+\frac{\alpha}{N}}_{C}\ket{r}_{S},
\end{equation}
where we have used the fact that
\begin{equation}
    e^{-i\alpha \hat{V}_\mathrm{int}(\hat{T})} = \int_\mathds{R}dt\, \ketbra{t}{t}\otimes e^{-i\alpha f(t)\hat{K}_{S}(t)} .
\end{equation}
Repeating the process, one finds that
\begin{equation}
    \left(e^{-i\frac{\alpha}{N}\hat{H}_{C}} e^{-i\frac{\alpha}{N} \hat{V}_\mathrm{int}(\hat{T})}\right)^{2} \ket{t}_{C}\ket{r}_{S} = e^{-i\frac{\alpha}{N}  f(t+\frac{\alpha}{N}) \hat{K}_{S}(t+\frac{\alpha}{N})} e^{-i\frac{\alpha}{N}  f(t) \hat{K}_{S}(t)} \ket{t+2\frac{\alpha}{N}}_{C}\ket{r}_{S} .
\end{equation}
and thus that the $N^\text{th}$ power satisfies
\begin{align}
    \left(e^{-i\frac{\alpha}{N}\hat{H}_{C}} e^{-i\frac{\alpha}{N} \hat{V}_\mathrm{int}(\hat{T})}\right)^{N} \ket{t}_{C}\ket{r}_{S} &= \prod_{n=0}^{N-1} e^{-i\frac{\alpha}{N}  f(t+\frac{n}{N} \alpha) \hat{K}_{S}(t+\frac{n}{N}\alpha)}  \ket{t+\alpha}_{C}\ket{r}_{S} \\
    &= e^{-i \alpha\hat{H}_{C}} \prod_{n=0}^{N-1} e^{-i\frac{\alpha}{N}  f(t+\frac{n}{N} \alpha) \hat{K}_{S}(t+\frac{n}{N}\alpha)}  \ket{t}_{C}\ket{r}_{S} \\
    &= e^{-i \alpha\hat{H}_{C}} \int_\mathds{R}dt'\, \ketbra{t'}{t'}\otimes \prod_{n=0}^{N-1} e^{-i\frac{\alpha}{N}  f(t'+\frac{n}{N} \alpha) \hat{K}_{S}(t'+\frac{n}{N}\alpha)}  \ket{t}_{C}\ket{r}_{S}, \label{eTrotterDeriv} 
\end{align}
where $\prod_{n=0}^{N-1}$ should be understood as successive left-multiplication.
Since Eq.~\eqref{eTrotterDeriv} holds for all $t$ and $r$, we can simply equate the operators themselves, i.e. 
\begin{equation} \label{ePreTrotterSuzuki}
    \left(e^{-i\frac{\alpha}{N}\hat{H}_{C}} e^{-i\frac{\alpha}{N} \hat{V}_\mathrm{int}(\hat{T})}\right)^{N} = e^{-i \alpha\hat{H}_{C}}  \int_\mathds{R}dt\, \ketbra{t}{t}\otimes \prod_{n=0}^{N-1} e^{-i\frac{\alpha}{N}  f(t+\frac{n}{N} \alpha) \hat{K}_{S}(t+\frac{n}{N}\alpha)} .
\end{equation}
Then, as $N\to\infty$ in Eq,~\eqref{eTrotter}, the time-dependent Trotter-Suzuki formula (see e.g.~\cite{sun2020trotterized}) applied to the product in Eq.~\eqref{ePreTrotterSuzuki} gives
\begin{equation}
    e^{-i\alpha \hat{C}_\mathrm{int}} = e^{-i \alpha\hat{H}_{C}}  \int_\mathds{R}dt\, \ketbra{t}{t}\otimes \mathcal{T} \exp\left[ -i \int_{t}^{t+\alpha}ds \, f(s) \hat{K}_{S}(s) \right] ,
\end{equation}
where $\mathcal{T}$ denotes the time-ordered exponential and we have performed a change of integration variables from $s\to s-t$ (which, being monotone increasing in $s$, does not affect the time-ordering). The time-ordered exponential can be written in terms of the Dyson series:
\begin{equation} \label{eDysonSer}
\begin{aligned}
    e^{-i\alpha\hat{C}_\mathrm{int}} = e^{-i\frac{\alpha}{N}\hat{H}_{C}} \int_\mathds{R}dt\,\ketbra{t}{t}\otimes\left\lbrace \mathds{1} - i \int_{0}^{\alpha}ds_{1} \, f(s_{1}+t) \hat{K}_{S}(s_{1}+t) \right. \\
    \left. - \int_{0}^{\alpha}ds_{2}\, \int_{0}^{s_{2}}ds_{1} \, f(s_{2}+t) f(s_{1}+t) \hat{K}_{S}(s_{2}+t) \hat{K}_{S}(s_{1}+t) + \ldots \right\rbrace .
\end{aligned}
\end{equation}
Note that, this expression can also be obtained by the methods presented in~\cite{Smith2019} by equating $\hat{H}_{\rm int}$ therein with Eq.~\eqref{eq:H_int} above (cf. Eqs.~(17)-(19) in~\cite{Smith2019}).

\subsection{Instantaneous interventions}

A perfectly well-timed intervention, such as $f(t)=\delta(t)$, is strictly speaking ill-defined, since the exponential of Eq.~\eqref{eAppConstr} will in that case lead to powers of $\delta(0)$, and therefore diverge. However, we may consider the case where $f(t)$ is arbitrarily well localized around some value, and we can approximate this situation by using the Dirac delta distribution in the Dyson series in Eq.~\eqref{eDysonSer}. For example, considering such a perfectly-timed intervention occurring at clock time $\bar{\tau}$, we have  $f(t)=\delta(t-\bar{\tau})$, and then Eq.~\eqref{eDysonSer} gives
\begin{equation}  \label{eInstIntDyson}
\begin{aligned}
    e^{-i\alpha\hat{C}_\mathrm{int}} = e^{-i\alpha\hat{H}_{C}} \int_\mathds{R}dt\,\ketbra{t}{t}\otimes\left\lbrace \mathds{1} - i \int_{0}^{\alpha}ds_{1} \, \delta(s_{1}+t-\bar{\tau}) \hat{K}_{S}(s_{1}+t) \right. \\
    \left. - \int_{0}^{\alpha}ds_{2} \int_{0}^{s_{2}}ds_{1} \, \delta(s_{2}+t-\bar{\tau})\delta(s_{1}+t-\bar{\tau}) \hat{K}_{S}(s_{2}+t) \hat{K}_{S}(s_{1}+t) + \ldots \right\rbrace .
\end{aligned}
\end{equation}
Considering the terms in the series individually, we have
\begin{equation}
    \int_{0}^{\alpha}ds_{1} \, \delta(s_{1}+t-\bar{\tau}) \hat{K}_{S}(s_{1}+t) = \begin{cases}
\hat{K}_{S}(\bar{\tau}) & \text{if } \bar{\tau} \in [t,t+\alpha],\\
0  & \text{otherwise},
\end{cases}
\end{equation}
and for the second-order term, noting that $\delta(s_{2}+t-\bar{\tau})\delta(s_{1}+t-\bar{\tau})=0$ if $s_{1}\neq s_{2}$, we find that
\begin{equation}
    \int_{0}^{\alpha}ds_{2} \int_{0}^{s_{2}}ds_{1} \, \delta(s_{2}+t-\bar{\tau})\delta(s_{1}+t-\bar{\tau}) \hat{K}_{S}(s_{2}+t) \hat{K}_{S}(s_{1}+t) = \begin{cases}
\hat{K}_{S}(\bar{\tau})^{2} & \text{if } \bar{\tau} \in [t,t+\alpha],\\
0  & \text{otherwise}.
\end{cases}
\end{equation}
and so on for each term, and thus Eq.~\eqref{eInstIntDyson} becomes
\begin{equation}
    e^{-i\alpha\hat{C}_\mathrm{int}} = e^{-i\alpha\hat{H}_{C}} \int_\mathds{R}dt\,\ketbra{t}{t}\otimes e^{-i\,\mathrm{rect}_{\alpha}(\bar{\tau}-t)\hat{K}_{S}(\bar{\tau})},
\end{equation}
where $\mathrm{rect}_{\alpha}(s)$ is a rectangular function defined such as 
\begin{equation}
	\mathrm{rect}_{\alpha}(s) \coloneqq \begin{cases}
1 & \text{if } s \in [0,\alpha],\\
0  & \text{otherwise.} \end{cases}
\end{equation}
Then
\begin{equation}
    e^{-i\alpha\hat{C}_\mathrm{int}} = e^{-i\alpha\hat{H}_{C}} \int_\mathds{R}dt\,\ketbra{t}{t}\otimes e^{-i\,\mathrm{rect}_{\alpha}(\bar{\tau}-t)\hat{K}_{S}(\bar{\tau})},
\end{equation}

We will now show how this can give a Page-Wootters picture with the usual evolution under $\hat{H}_C +\hat{H}_S$, interspersed with an intervention occurring at time $\bar{\tau}$. Recalling that an arbitrary physical state can be obtained via a kinematical state of the form $\ket{t=0}_{C}\ket{\phi_{0}}_{S}$ (see Sec.~\ref{ssec:singelclockPW}) we consider the action of $e^{-i\alpha\hat{C}_\mathrm{int}}$ on such a state. Assuming for simplicity that $\bar{\tau}>0$,
\begin{equation}
    e^{-i\alpha\hat{C}_\mathrm{int}}\ket{t=0}_{C}\ket{\phi_{0}}_{S} =
    \begin{cases}
        e^{-i\alpha\hat{H}_{C}}\ket{t=0}_{C}\ket{\phi_{0}}_{S},  &\text{if } \alpha<\bar{\tau}\\
        \left(e^{-i\alpha\hat{H}_{C}}\otimes e^{-i\hat{K}_{S}(\bar{\tau})}\right)\ket{t=0}_{C}\,\ket{\phi_{0}}_{S},  &\text{if } \alpha \geq \bar{\tau}.
    \end{cases}
\end{equation}
Recalling that the interaction-picture physical state corresonding to $\ket{t=0}_{C}\ket{\phi_{0}}_{S}$ is given by $\delta(\hat{C}_\text{int})\ket{t=0}_{C}\ket{\phi_{0}}_{S}$, and that a physical state can be obtained from the interaction picture via $\kket{\Psi} \equiv U_\text{int}^{-1}\kket{\Psi_\text{int}}$, we have
\begin{equation}
\begin{aligned}
    \kket{\Psi} &\equiv U_\text{int}^{-1}\,\delta(\hat{C}_\text{int})\ket{t=0}_{C}\ket{\phi_{0}}_{S} = e^{-i\hat{T}\otimes\hat{H}_{S}} \frac{1}{2\pi}\int_\mathds{R} d\alpha \, e^{-i\alpha\hat{H}_{C}}\ket{t=0}_{C}\ket{\phi_{0}}_{S} \\
    & = e^{-i\hat{T}\otimes\hat{H}_{S}} \left\lbrace \frac{1}{2\pi} \int_{-\infty}^{\bar{\tau}} dt \ket{t}_{C}\ket{\phi_{0}}_{S} + \frac{1}{2\pi} \int_{\bar{\tau}}^{\infty} dt \ket{t}_{C}\,e^{-i\hat{K}_{S}(\bar{\tau})}\ket{\phi_{0}}_{S} \right\rbrace .
\end{aligned}
\end{equation}
Noting that, as in the usual interaction picture, $\hat{K}_{S}(\bar{\tau})^{n}=e^{i\bar{\tau}\hat{H}_{S}}\hat{K}_{S}^{n}e^{-i\bar{\tau}\hat{H}_{S}}$, one can write the physical state $\kket{\Psi}$ above as
\begin{equation}
    \kket{\Psi} =  \frac{1}{2\pi} \int_{-\infty}^{\bar{\tau}} dt \ket{t}_{C} \,e^{-it\hat{H}_{S}}\ket{\phi_{0}}_{S} + \frac{1}{2\pi} \int_{\bar{\tau}}^{\infty} dt \ket{t}_{C}\,e^{-i(t-\bar{\tau})\hat{H}_{S}}e^{-i\hat{K}_{S}}\,e^{-i\bar{\tau}\hat{H}_{S}}\ket{\phi_{0}}_{S} .
\label{eq:state_gio}
\end{equation}
In other words, the reduced physical (i.e. Page-Wootters) state $\ket{\psi_{S\vert C}(t)}$ corresponds to relational evolution under $H_{S}$ for $-\infty<t<\bar{\tau}$ consistent with the ``initial condition'' $\ket{\psi_{S\vert C}(0)}=\ket{\phi_{0}}_{S}$, and then subsequent relational evolution under $H_{S}$ consistent with the ``initial condition'' $\ket{\psi_{S\vert C}(0)}=\ket{\phi_{0}}_S$ as well as with the unitary intervention $e^{-i\hat{K}_{S}}$ having occurred at $t=\bar{\tau}$. The state in Eq.~\eqref{eq:state_gio} has the same structure as the one in Eq.~(44) of~\cite{giovannetti2015quantum}, where the authors consider the particular case of measurements.
By replacing the interaction term $\hat{f}(\hat{T})\otimes K_{S}$ with multiple interactions, one obtains a corresponding segment-wise evolution, as employed in Sec.~\ref{sec:OperationalCausality&PW} of the main text.
\end{appendix}

\end{document}